\documentclass[sigconf]{acmart}

\AtBeginDocument{%
  \providecommand\BibTeX{{%
    \normalfont B\kern-0.5em{\scshape i\kern-0.25em b}\kern-0.8em\TeX}}}

\usepackage{hyperref}
\usepackage{algorithm}
\usepackage{algorithmic}
\usepackage{subfigure}
\usepackage{amsthm}
\usepackage{color}
\usepackage{makecell}
\usepackage{multirow}
\usepackage{bm}
\usepackage{amsfonts}
\usepackage{url}
\usepackage{subfigure}
\usepackage{balance}

\copyrightyear{2022}
\acmYear{2022}
\setcopyright{acmcopyright}
\acmConference[SIGIR '22] {Proceedings of the 45th International ACM SIGIR Conference on Research and Development in Information Retrieval}{July 11--15, 2022}{Madrid, Spain.}
\acmBooktitle{Proceedings of the 45th International ACM SIGIR Conference on Research and Development in Information Retrieval (SIGIR '22), July 11--15, 2022, Madrid, Spain}
\acmPrice{15.00}
\acmISBN{978-1-4503-8732-3/22/07}
\acmDOI{10.1145/3477495.3531974}

\begin{document}
\fancyhead{}

\title{Explainable Legal Case Matching via Inverse Optimal Transport-based Rationale Extraction}

\author{Weijie Yu}
\affiliation{
\institution{School of Information\\Renmin University of China}
  \city{Beijing}\country{China}
}
\email{yuweijie@ruc.edu.cn}

\author{Zhongxiang Sun, Jun Xu}
\authornote{Jun Xu is the corresponding author. Work partially done at Beijing Key Laboratory of Big Data Management and Analysis Methods.
}
\affiliation{%
  \institution{GSAI, Renmin University of China}
  \city{Beijing}\country{China}
  }
\email{jeryi.sunzx01@gmail.com}
\email{junxu@ruc.edu.cn}


\author{Zhenhua Dong}
\affiliation{%
  \institution{Noah’s Ark Lab, Huawei}
  \city{Shenzhen}
  \country{China}
  }
\email{dongzhenhua@huawei.com}


\author{Xu Chen}
\affiliation{%
  \institution{GSAI, Renmin University of China}
  \city{Beijing}
  \country{China}
  }
\email{xu.chen@ruc.edu.cn}

\author{Hongteng Xu}
\affiliation{%
  \institution{GSAI, Renmin University of China}
  \city{Beijing}
  \country{China}
  }
\email{hongtengxu@ruc.edu.cn}

\author{Ji-Rong Wen}
\affiliation{%
  \institution{GSAI, Renmin University of China}
  \city{Beijing}
  \country{China}
  }
\email{jrwen@ruc.edu.cn}








\begin{abstract}
As an essential operation of legal retrieval, legal case matching plays a central role in intelligent legal systems.
This task has a high demand on the explainability of matching results because of its critical impacts on downstream applications --- the matched legal cases may provide supportive evidence for the judgments of target cases and thus influence the fairness and justice of legal decisions. 
Focusing on this challenging task, we propose a novel and explainable method, namely \textit{IOT-Match}, with the help of computational optimal transport, which formulates the legal case matching problem as an inverse optimal transport (IOT) problem. 
Different from most existing methods, which merely focus on the sentence-level semantic similarity between legal cases, our IOT-Match learns to extract rationales from paired legal cases based on both semantics and legal characteristics of their sentences. 
The extracted rationales are further applied to generate faithful explanations and conduct matching.
Moreover, the proposed IOT-Match is robust to the alignment label insufficiency issue commonly in practical legal case matching tasks, which is suitable for both supervised and semi-supervised learning paradigms. 
To demonstrate the superiority of our IOT-Match method and construct a benchmark of explainable legal case matching task, we not only extend the well-known Challenge of AI in Law (CAIL) dataset but also build a new Explainable Legal cAse Matching (ELAM) dataset, which contains lots of legal cases with detailed and explainable annotations.
Experiments on these two datasets show that our IOT-Match outperforms state-of-the-art methods consistently on matching prediction, rationale extraction, and explanation generation.
\end{abstract}




\maketitle
\section{Introduction}
\label{sec:intro}

Legal case matching aims at identifying relations between paired legal cases, which is a key task of legal retrieval.
This task has a high demand on the explainability of matching results because of its critical impacts on legal justice. 
In particular, the matching results may indicate significant evidence or information, which influences the incentives of decision-makers in the common law system and provides the basis for legal reasoning in the civil law system.


To achieve this aim, many efforts have been made, including the early attempts that are based on rule-based strategies~\cite{zeng2005knowledge,saravanan2009improving,bench2012history} and the recent learning-based methods like the Precedent Citation Network (PCNet)~\cite{kumar2011similarity} and BERT-based methods~\cite{shao2020bert,xiao2021lawformer}.
Although these methods have achieved encouraging performance, they often suffer from the following challenges on providing plausible and faithful explanations associated with matching results.
Firstly, legal cases are long-form documents with complicated contents in general, in which only the rationales representing the legal characteristics should support matching results and their explanations.
However, existing methods tend to overlook the striking different roles between the rationales and other sentences~\cite{shao2020bert,chalkidis2020legal,xiao2021lawformer}.
Secondly, ideal explanations of legal case matching results are expected to offer reasons for one side and to rebut arguments for the other side~\cite{atkinson2020explanation}, but existing methods often fail to distinguish the pro rationales and con rationales that respectively support the matching and mismatching decisions~\cite{paranjape2020information,sha2021learning,luu2021explaining}. 
Moreover, the ground-truth labels for explanations (e.g., aligned rationales across different cases) are expensive, sparse, and usually biased (e.g., only limited number of positive pairs are labeled correctly while lots of false negative pairs exist). 
As a result, learning based on such labeled data often leads to sub-optimal matching results and unreliable explanations.

\begin{figure}[t]
    \centering
    \includegraphics[width=\linewidth]{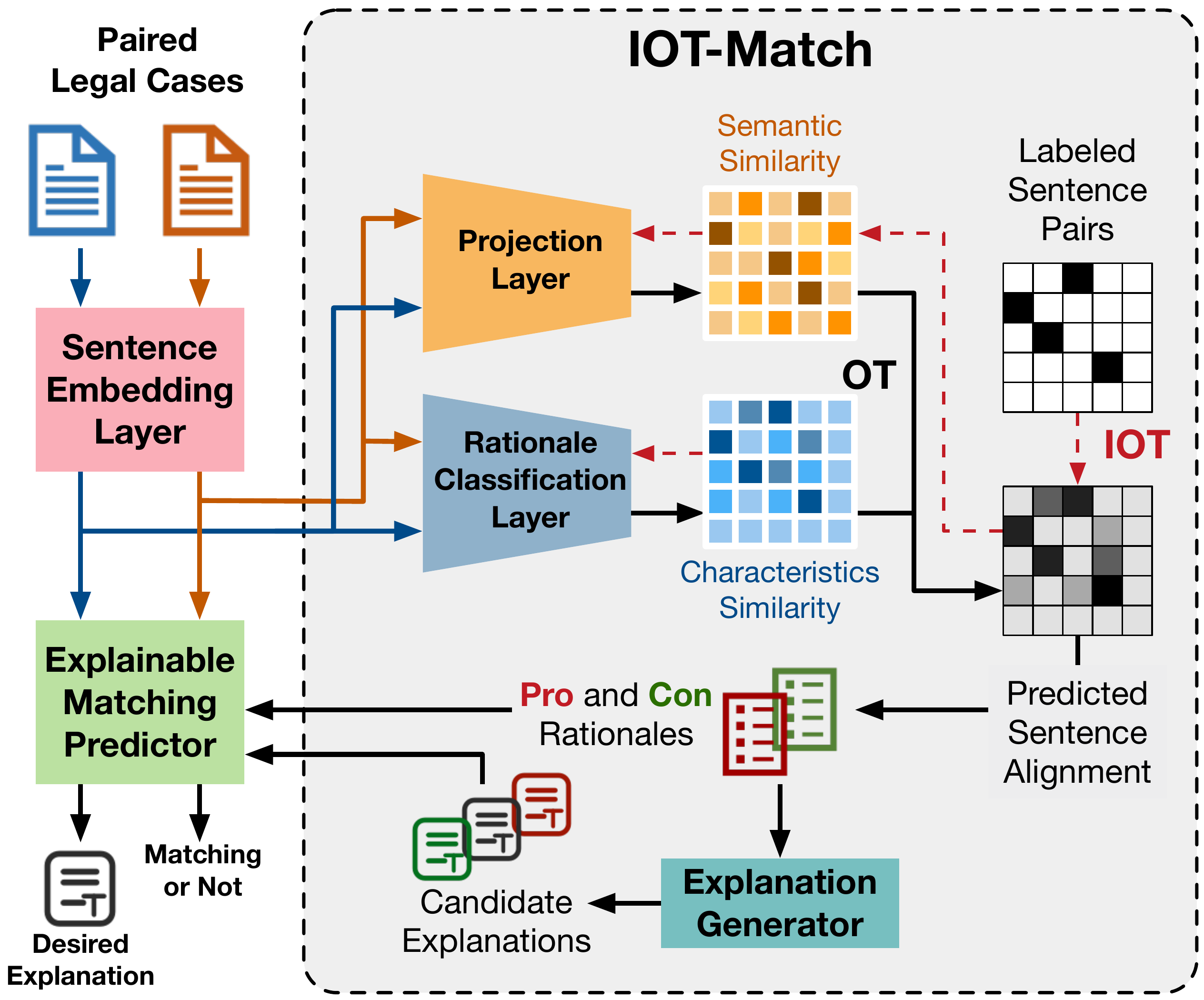}
    \caption{The architecture of our model IOT-Match. 
    Note that the red dotted arrows indicate the back-propagation achieved by inverse optimal transport, which are used only in the training phase.}
    \label{fig:architecture}
\end{figure}

Facing the above challenges, in this paper, we propose a novel inverse optimal transport~\cite{dupuy2016estimating,li2019learning} (IOT)-based model, called IOT-Match, to extract rationales for explainable legal cases matching. 
As illustrated in Figure~\ref{fig:architecture}, our IOT-Match formulates the extraction and alignment of pro and con rationales as an optimal transport (OT) problem, in which the identified rationales and their alignments are derived from the transport plan of the OT solution. 
The optimal transport is guided by a learnable affinity matrix that reflects both semantics and legal characteristic relations between cross-case sentences, and the affinity matrix is learned by the IOT process, which corresponds to solving a bi-level optimization problem. 
In this way, IOT-Match learns to extract the pro and con rationales directly. 
To apply the proposed model to real legal case matching applications and following the practices in~\citep{kumar2020nile,zhao2021lirex}, the extracted rationales from the paired legal cases are then fed to a pre-trained language model to generate label-specific natural languages explanations that stand for the pro and con reasons of matching. 
For filtering out the noise sentences and weighing the pro and con reasons, the final matching results are made based on the extracted rationales and the generated label-specific explanations. 

Besides proposing an explainable legal case matching method, we construct a new dataset called \textbf{E}xplanable \textbf{L}egal c\textbf{A}se \textbf{M}atching (ELAM). 
To be best of our knowledge, our ELAM is the first legal case matching dataset which provides matching labels for legal case pairs and detailed annotations, including rationales, alignments, and natural language-based explanations for matching labels.

In summary, our contributions include the following three folds: 
(1) We propose a novel model, namely IOT-Match, to extract rationales and generate natural language-based explanations for legal case matching.
To the best of our knowledge, IOT-Match is the first explainable model for legal case matching. 
(2) We construct a new large-scale dataset ELAM which facilitates future research on explainable legal case matching.
(3) Experimental results indicate IOT-Match not only achieves state-of-the-art performance in legal case matching but also produces plausible and faithful explanations for its matching prediction.

\section{Related Work}
\subsection{Legal Case Matching}

Conventional legal case matching methods highly depend on expert knowledge~\cite{bench2012history}, e.g., the decomposition of legal issues~\cite{zeng2005knowledge} and the ontological framework of the problem~\cite{saravanan2009improving}.
In recent years, learning-based legal case matching strategy has shown advantages in exploring the semantics of legal cases, which can be roughly categorized into network-based methods~\cite{kumar2011similarity,monroy2013link,minocha2015finding,bhattacharya2020hier} and text-based methods~\cite{shao2020bert,xiao2021lawformer}. 
The network-based methods construct a Precedent Citation Network (PCNet), in which the vertices are legal cases and directed edges indicate the citations of source cases used by target cases. 
Based on PCNet, \citet{kumar2011similarity} used the Jaccard similarity index between the sets of precedent citations to infer the similarity of two legal cases.
~\citet{minocha2015finding} used whether the sets of precedent citations occurs in the same cluster to measures to what extent the two cases are similar.
~\citet{bhattacharya2020hier} proposed Hier-SPCNet to capture all domain information inherent in both statutes and precedents. 
The text-based methods rely on the textual content of the cases and measure the similarity of two legal cases based on their semantics. 
\citet{shao2020bert} proposed BERT-PLI to break a case into paragraphs and model the interactions between the paragraphs. 
It first adopted BERT to encode each paragraph in two legal cases, then applied max-pooling to capture their matching signal, and finally, used a recurrent neural network (RNN) with an attention mechanism to predict their matching score. 
Similarly,~\citet{bhattacharya2020methods} proposed to segment two legal cases into paragraphs and aggregate the paragraph-level similarity.
Inspired by the success of pre-trained language models in the generic domain,~\citet{xiao2021lawformer} pre-trained a Longformer-based language model with tens of millions of criminal and civil case documents. 
Although these studies effectively improve the performance, they often have difficulties on explaining their predictions, which limits their practical applications~\cite{bibal2021legal}.
\subsection{Explainable AI in the Legal Domain}
Recently, researchers have made some efforts to achieve explainable AI models in various applications of the legal domain~\cite{doshi2017accountability}.
In the task of legal judgment prediction, \citet{ye2018interpretable} formalized the court view generation problem as a label-conditioned Seq2Seq task and generated court views based on fact descriptions and charges.
~\citet{jiang2018interpretable} proposed a neural based system to jointly extract readable rationales and elevate charge prediction accuracy by a rationale augmentation mechanism.
~\citet{liu2021interpretable} proposed the Joint Prediction and Generation Model (JPGM) to predict charges and court views. 
JPGM generated charge-discriminative information and used a coarse-to-fine classifier to effectively deal with the confusions within charges. 
In addition, JPGM explicitly modeled the interdependence between charges and court views.
In the task of legal question answering, ~\citet{zhong2020iteratively} proposed to first detect elements of fact descriptions by iteratively asking questions about pre-defined charge-specific principles and then used the detected elements for prediction. 
Different from the above work, we focus on the legal case matching task, extracting rationales and generating natural language-based explanations to support matching results. 

\section{Proposed IOT-Match Method}
\subsection{Problem Statement}\label{sec:formulation}
To conduct explainable legal case matching, we are given a set of labeled data tuples $\mathcal{D} = \{(X, Y, \mathbf{r}^X, \mathbf{r}^Y, \mathbf{\hat{A}}, z, e)\}$. 
For each tuple $(X, Y, \mathbf{r}^X, \mathbf{r}^Y, \mathbf{\hat{A}}, z, e)$ in the dataset, its elements include: 1) a pair of legal cases $X\in\mathcal{X}$ and $Y\in\mathcal{Y}$, where $\mathcal{X}$ and $\mathcal{Y}$ represent the sets of source and target legal cases; 
2) the rationale labels of the paired cases, denoted as $\mathbf{r}^X$ and $\mathbf{r}^Y$, respectively; 
3) a binary alignment matrix $\mathbf{\hat{A}}$ indicating the semantic relation between rationales of
$X$ and rationales of $Y$;
and 4) the matching label $z$ and the set of sentences (denoted as $e$) explaining the reasons for $z$.

\begin{figure*}
    \centering
    \subfigure[case $X$ (16 sentences)]{
    \includegraphics[width=0.372\linewidth]{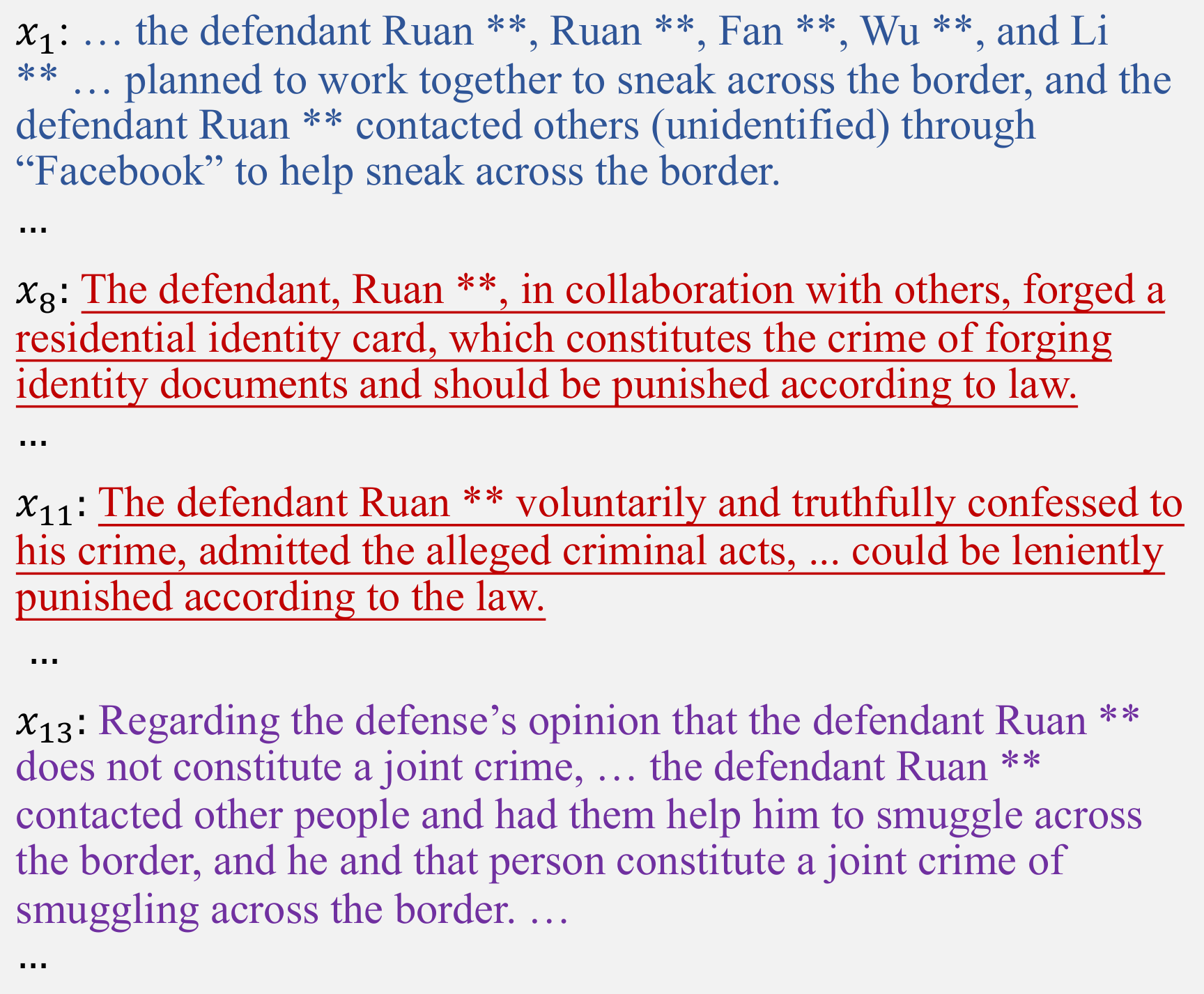}
    \label{fig:exmaple-a}
    }
    \subfigure[case $Y$ (8 sentences)]{
    \includegraphics[width=0.28\linewidth]{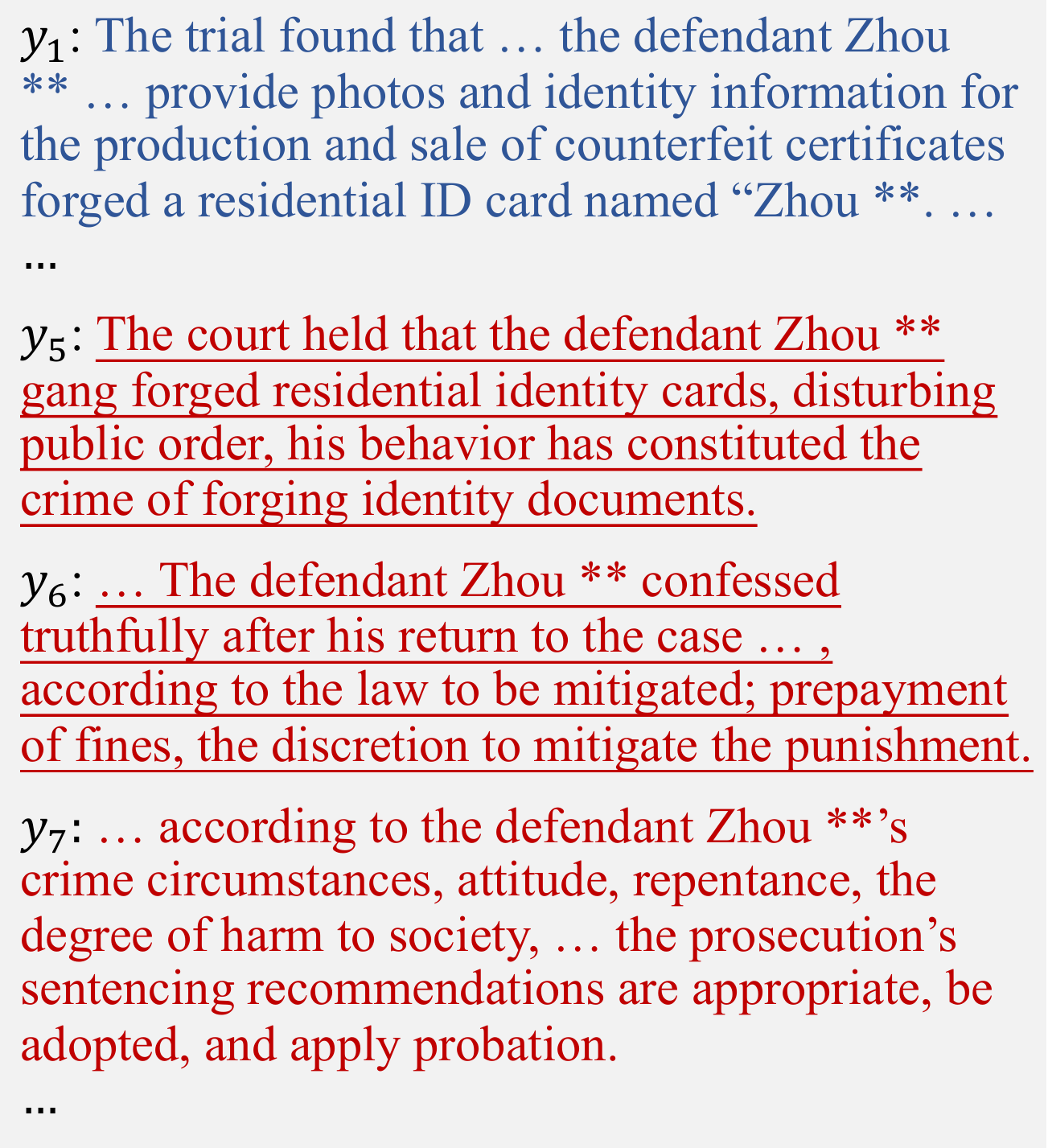}
    \label{fig:exmaple-b}
    }
    \subfigure[human annotations: $\mathbf{r}^X, \mathbf{r}^Y, \hat{\mathbf{A}}, z$, and $e$]{
    \includegraphics[width=0.285\linewidth]{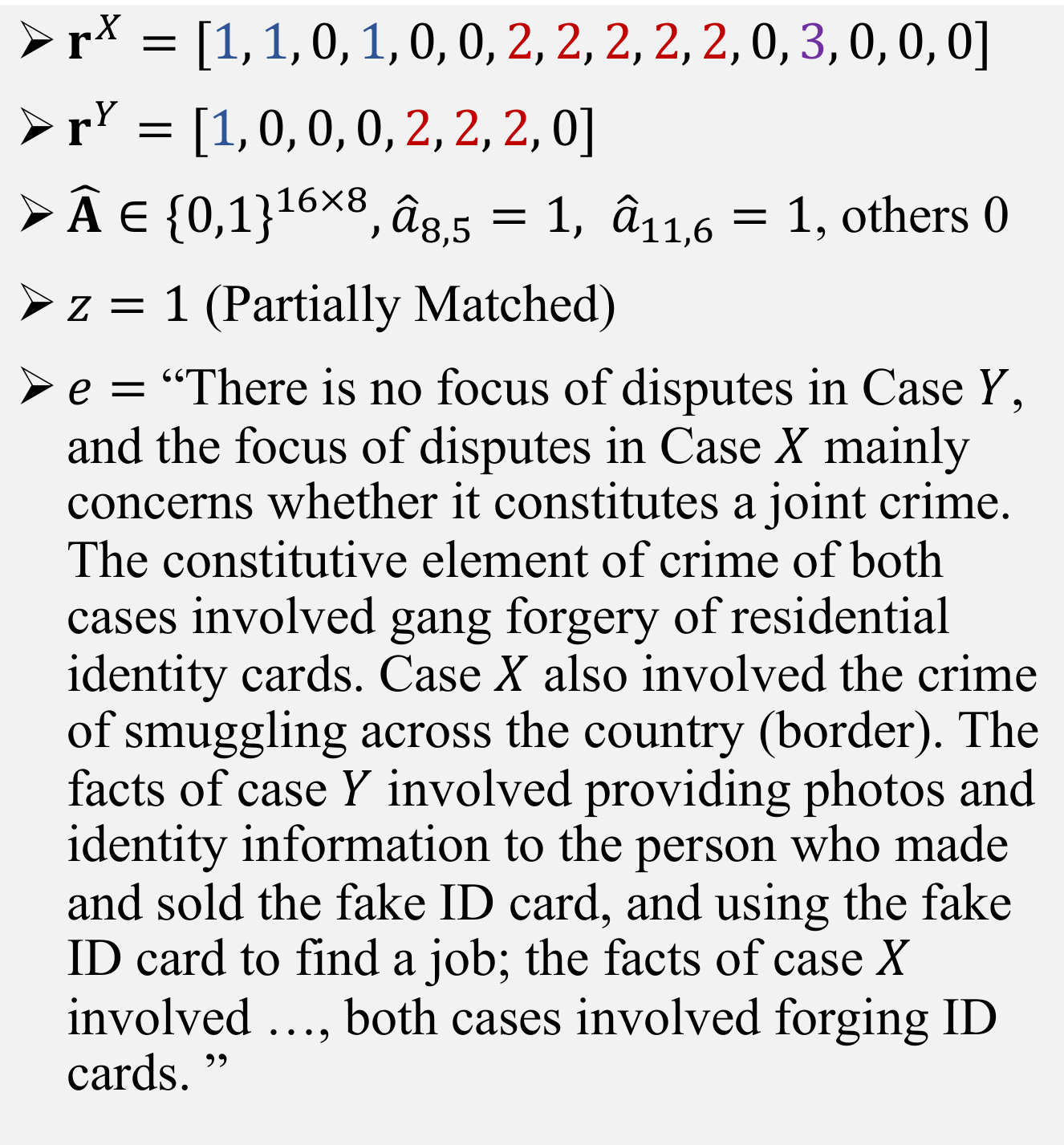}
    \label{fig:exmaple-c}
    }
    \caption{A labeled legal case pair (translated from Chinese). Blue, red, and purple denote rationale labels of $r_s=1,2$ and $3$, respectively. The underlined rationales are aligned. Note some sentences are omitted for the convenience of representation.
    }
    \label{fig:exmaple}
\end{figure*}

In practice, we represent each legal case as a set of sentence-level embeddings, i.e., $X = \{x_m\}_{m=1}^{M}$ and $Y = \{y_n\}_{n=1}^{N}$, where $x_m$ ($y_n$) denotes the embedding of the $m$-th ($n$-th) sentence in $X$ ($Y$). 
Typically, each embedding can be calculated by using the output of at the [CLS] token of a BERT model pre-trained on a Chinese legal case corpus.~\footnote{Corpus available at \url{https://github.com/thunlp/OpenCLaP}. Note that IOT-Match is applicable to the corpus of other languages with the corresponding embeddings.}
The rationale labels are associated with the sentence embeddings, i.e., $\mathbf{r}^X = \{r_{x_m}\}_{m=1}^{M}$ and $\mathbf{r}^Y = \{r_{y_n}\}_{n=1}^{N}$, where the rationale label of a sentence $s$ is designed following~\cite{ma2021lecard}:
\begin{eqnarray}\label{eq:rlabel}
r_s=
\begin{cases}
0 &\text{$s$ is not a rationale},\\
1 &\text{$s$ is a key circumstance},\\
2 &\text{$s$ is a constitutive element of crime},\\
3 &\text{$s$ is a focus of disputes}.
\end{cases}
\end{eqnarray}
The remaining elements, i.e., $\mathbf{\hat{A}}=[\hat{a}_{mn}]\in \{0, 1\}^{M\times N}$, $z\in \{0, 1, 2\}$, and $e$, are annotated manually, where 
\begin{eqnarray}\label{eq:zlabel}
\hat{a}_{mn}=
\begin{cases}
0 & r_{x_m}\neq r_{y_n},\\
1 & r_{x_m}=r_{y_n}~\&~x_m\cong y_n,
\end{cases}
z=
\begin{cases}
0 & \text{Mismatched $(X, Y)$},\\
1 & \text{Partially matched},\\
2 & \text{Matched},\\
\end{cases}
\end{eqnarray}
where $x_m\cong y_n$ means the sentences corresponding to $x_m$ and $y_n$ are semantically-similar. 
$\hat{a}_{mn}=1$ means aligned rationales while $\hat{a}_{mn}=0$ means misaligned rationales. 
They provide pro and con evidence for matching prediction, respectively. Figure~\ref{fig:exmaple} shows an example of human labeled explainable legal case pair.

The proposed explainable legal case matching aims at learning the following three modules: 
1) $f_1$ extracts aligned and misaligned rationales from the paired legal cases (\textbf{Sec.~\ref{sec:stage1}}); 
2) $f_2$ generates candidate explanations based on the rationales extracted by $f_1$ (\textbf{Sec.~\ref{sec:generation}}); 
and 3) $f_3$ predicts the final matching label based on the extracted rationales and generated explanations (\textbf{Sec.~\ref{sec:matching}}).
\subsection{The Principle of Our Method}\label{sec:ot}
As the key of the above three modules, $f_1$ is learned to fit the given alignment matrices $\mathbf{\hat{A}}$'s.
As aforementioned, however, the alignment matrices are manually labeled, which are often very sparse and thus contain many false negative elements. 
To learn our model robustly from such imperfect data, we develop a novel learning paradigm from the viewpoint of optimal transport (OT). 
Essentially, optimal transport~\cite{villani2009optimal,peyre2019computational} defines a distance between probability distributions, which has been widely used in many machine learning tasks, such as
point cloud alignment~\cite{alvarez2019towards,alvarez2018gromov}, graph matching~\cite{xu2019gromov,xu2019scalable,chen2020graph}, data clustering~\cite{xu2018distilled,chakraborty2020hierarchical}, and sequence representations learning~\cite{chen2018improving,yu-etal-2020-wasserstein,yu2022distribution}. 
In our scenario, following existing studies~\cite{chen2018improving,chen2020graph,xu2019gromov,alvarez2018gromov}, given two sets of sentence embeddings (e.g., the $X$ and $Y$ mentioned above),  we assume their empirical distributions to be uniform, i.e., $\bm{\mu}=\frac{1}{M}\mathbf{1}_M$ and $\bm{\nu}=\frac{1}{N}\mathbf{1}_N$, where $\mathbf{1}_D$ represents the $D$-dimensional all-one vector, and accordingly, compute the  optimal transport distance between them in a discrete format~\cite{cuturi2013sinkhorn}:
\begin{equation}\label{eq:dot}
\begin{aligned}
\mathbf{{A}}^*&=\mathop{\arg\sideset{}{_{\mathbf{A}\in\Pi(\bm{\mu},\bm{\nu})}}\min}\mathbb{E}_{m,n\sim \mathbf{A}}[c(x_m,y_n)]\\
&=\mathop{\arg\sideset{}{_{\mathbf{A}\in\Pi(\bm{\mu},\bm{\nu})}}\min}\sideset{}{_{m=1}^M}\sum\sideset{}{_{n=1}^N}\sum a_{mn}\cdot c(x_m,y_n),
\end{aligned}
\end{equation}
where $\mathbf{A}\in\Pi(\bm{\mu},\bm{\nu})=\{\mathbf{A}\in\mathbb{R}_+^{M\times N}|\mathbf{A}\mathbf{1}_N=\bm{\mu},\mathbf{A}^\top\mathbf{1}_M=\bm{\nu}\}$, which represents an arbitrary joint distribution of the sentences with marginals $\bm{\mu}$ and $\bm{\nu}$. 
$\mathbf{C}=[c(x_m,y_n)]\in\mathbb{R}^{M\times N}$ is a sentence-level affinity matrix, whose element $c(x_m, y_n)$ measures the discrepancy between the two sentences based on their embeddings. 
As shown in Eq.~\eqref{eq:dot}, the optimal transport distance actually corresponds to the minimum expectation of the sentence-level discrepancy, in which the optimal joint distribution $\mathbf{{A}}^*$ is called ``optimal transport''. 
The optimal transport matrix provides a soft matching between the sentence sets in a probabilistic way --- the element of the optimal transport, $i.e.$, ${a}_{mn}^*$, indicates the probability of the coherency of $x_m$ and $y_n$, which provides the evidence for their matching.

Note that Eq.~\eqref{eq:dot} is a linear programming. 
To accelerate the computation of optimal transport, in practice we often introduce an entropic regularizer into Eq.~\eqref{eq:dot}, which leads to an entropic optimal transport problem~\cite{cuturi2013sinkhorn}:
\begin{eqnarray}\label{eq:eot}
\mathbf{{A}}^*=\mathop{\arg\sideset{}{_{\mathbf{A}\in\Pi(\bm{\mu},\bm{\nu})}}\min}\langle\mathbf{A}, \mathbf{C}\rangle + \gamma \langle \mathbf{A},\log\mathbf{A}\rangle.
\end{eqnarray}
Here, $\langle\mathbf{A}, \mathbf{C}\rangle=\mathrm{Tr}(\mathbf{A}^\top\mathbf{C})$  represents the Frobenius dot-product, which represents the objective function of Eq.~\eqref{eq:dot} in a matrix format.
$\langle \mathbf{A},\log\mathbf{A}\rangle=\sum_{m,n}a_{mn}\log a_{mn}$ is the proposed entropic regularizer. 
The entropic optimal transport in Eq.~\eqref{eq:eot} is a strictly convex problem, which can be solved by the Sinkhorn scaling algorithm efficiently~\cite{cuturi2013sinkhorn}.

When learning the rationale extraction module $f_1$ in the above OT framework, the critical learning tasks become: 1) learning the affinity matrix $\mathbf{C}$ to optimize the guidance to the computation of the optimal transport $\mathbf{{A}}^*$; and 2) fitting the optimal transport $\mathbf{{A}}^*$ robustly to the manually-labeled (noisy) alignment matrix $\mathbf{\hat{A}}$. 
In this work, we solve these two tasks jointly by solving the following inverse optimal transport (IOT) problem.

\subsection{IOT-based Rationale Extraction}\label{sec:stage1} 
According to the analysis above, we need to learn both the affinity matrix and the optimal transport based on the sentence embeddings ($X$ and $Y$) and their annotated alignment matrix $\mathbf{\hat{A}}$, which leads to a so-called inverse optimal transport (IOT) problem~\cite{dupuy2016estimating,li2019learning}: 
\begin{equation}\label{eq:IOT4Legal}
\begin{aligned}
&\mathbf{C}^*=\mathop{\arg\sideset{}{_{\mathbf{C}\in\mathbb{R}^{M\times N}}}\min} \text{KL}(\mathbf{\hat{A}}\|\mathbf{{A}}^*(\mathbf{C})),\\
&s.t.~\mathbf{{A}}^*(\mathbf{C})=\mathop{\arg\sideset{}{_{\mathbf{A}\in\Pi(\bm{\mu},\bm{\nu})}}\min}\langle\mathbf{A}, \mathbf{C}\rangle + \gamma \langle \mathbf{A},\log\mathbf{A}\rangle.
\end{aligned}
\end{equation}
This problem is a typical bi-level optimization problem, in which the affinity matrix $\mathbf{C}$ is the upper-level variable while the optimal transport $\mathbf{{A}}$ is the lower-level variable. 
The upper-level problem minimizes the KL divergence between $\mathbf{\hat{A}}$ and $\mathbf{{A}}^*$, i.e., $\text{KL}(\mathbf{\hat{A}}\|\mathbf{{A}}^*)=\sum_{m,n}\hat{a}_{mn}\log\frac{{a}_{mn}^*}{\hat{a}_{mn}}$, which corresponds to the cross-entropy loss. 
The optimal transport $\mathbf{{A}}^*$ is a function of the affinity matrix, $i.e.$, $\mathbf{{A}}^*(\mathbf{C})$, whose optimization corresponds to the lower-level problem given $\mathbf{C}$.

Solving the IOT problem in Eq.~\eqref{eq:IOT4Legal} provides us a robust method to learn the rationale extraction module. 
Specifically, on the one hand, the upper-level problem fits the optimal transport to the limited and noisy alignment matrix under the constraint provided by the lower-level optimal transport problem, which suppresses the risk of over-fitting greatly.
On the other hand, the lower-level problem provides us with an optimal transport matrix to indicate the aligned rationales, which is determined by the optimized affinity matrix and thus reveals sentence-level similarity between the paired legal cases. 
As a result, the optimal transport $\mathbf{A}^*$ derived from the optimal affinity matrix $\mathbf{C}^*$ represents the global alignment between rationales of a legal case pair.
Accordingly, we can extract pro (aligned) and con (misaligned) rationales by setting a threshold $\tau$, i.e., $x_m$ and $y_n$ are selected as pro rationales if $a_{mn}^*\ge\tau$, otherwise, are selected as con rationales. 

Note that because the lower-level problem is strictly convex, this IOT problem can be solved efficiently by alternating optimization. 
Given current affinity matrix $\mathbf{C}$, we can optimize $\mathbf{A}$ via the Sinkhorn scaling algorithm, and then optimize $\mathbf{C}$ via stochastic gradient descent based on fixed $\mathbf{A}$. 

We parameterize the affinity matrix $\mathbf{C}$ by a neural network, which takes paired sentence embeddings as its input and output their discrepancies according to their legal characteristics and semantics jointly. 
As illustrated in Figure~\ref{fig:cost}, we model $\mathbf{C}$ as the combination of a rationale characteristic matrix $\mathbf{C}^r$ and a semantic matrix $\mathbf{C}^s$:
\begin{equation}
    \mathbf{C} = \epsilon\mathbf{C}^r+\mathbf{C}^s,
\end{equation}
where $\epsilon$ is a negative hyper-parameter to encourage the alignment of those sentence pairs which have the same rationale label by significantly reducing the transport between them. The $\mathbf{C}^s$ and $\mathbf{C}^r$ are constructed by the following steps.

\subsubsection{Construction of the Semantic Matrix $\mathbf{C}^s$}
\begin{figure}[t]
    \centering
    \includegraphics[width=0.9\linewidth]{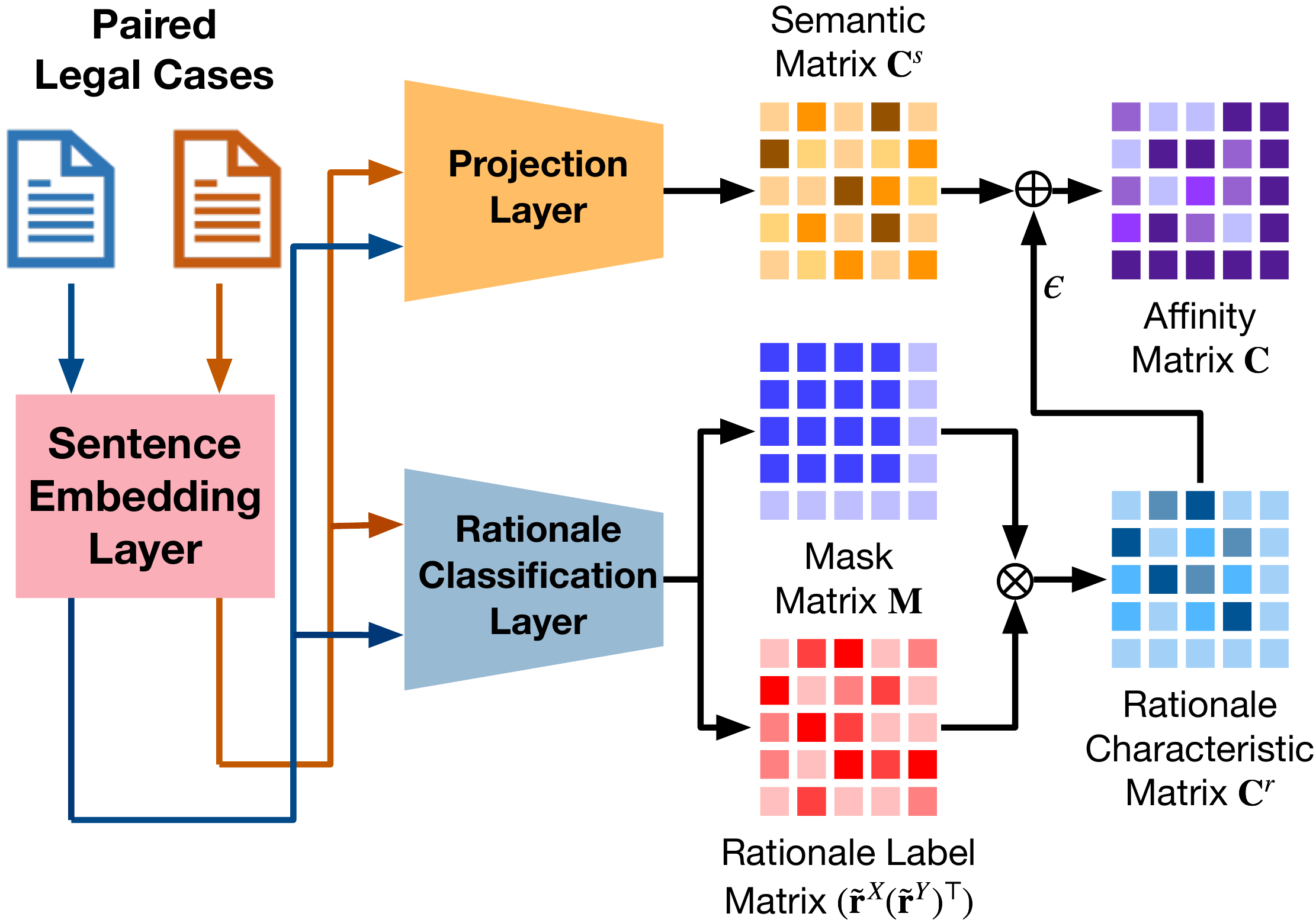}
    \caption{Illustration of the affinity matrix construction.}
    \label{fig:cost}
\end{figure}

IOT-Match constructs the semantic matrix $\mathbf{C}^s\in\mathbb{R}^{M\times N}$ to indicate the semantic distance between a cross-case sentence pair, i.e., $\mathbf{C}^s=\mathrm{dis}(\mathbf{\bar s}^X,\mathbf{\bar s}^Y)$, where function `dis' measures the semantic distance of two sentence embeddings (e.g., Euclidean distance), $\mathbf{\bar s}_{X}$ and $\mathbf{\bar s}_{Y}$ respectively represent the contextual sentence embedding of legal case $X$ and $Y$, which is obtained from a trainable two-layer MLP (projection layer in Figure~\ref{fig:cost}) on the frozen sentence embeddings $X$ and $Y$.

\subsubsection{Construction of the Rationale Characteristic Matrix $\mathbf{C}^r$}
The rationale characteristic matrix $\mathbf{C}^r$ indicates the rationales having the same legal characteristics, and the legal characteristics is categorized according to the rationale labels shown in Eq.~\eqref{eq:rlabel}.
Taking two legal cases $X$ and $Y$ as the inputs, our IOT-Match predicts the rationale labels of their sentences, denoted as $\mathbf{\hat{r}}^X=\{\hat{r}_{x_m}\}_{m=1}^{M}$ and $\mathbf{\hat{r}}^Y=\{\hat{r}_{y_n}\}_{n=1}^{N}$, respectively, which is achieved by solving a sentence-level multi-class classification problem. 
Formally, given a legal case $X$, our IOT-Match would identify the legal characteristics of each sentence embedding $x_m$ in $X$ by calculating a probabilistic distribution over the four classes shown in Eq.~\eqref{eq:rlabel}:
\begin{equation}
\label{eq:argmax}
    \hat{r}_{x_m} =\mathop{\arg\sideset{}{_{k\in\{0,\cdots 3\}}}\max} P(r = k|x_m),
\end{equation}
where $\{P(r=k|x_m)\}_{k=0}^{3}$ represent the distribution of the rationale labels conditioned on the sentence embedding $x_m$. 
In this work, we parameterize the distribution as follows:
\begin{align}
\label{eq:ClassDistr}
\{P(r=k|x_m)\}_{k=0}^3
=\text{softmax}(\mathbf{W} \mathbf{s}_{x_m}^{(L)} + \mathbf{b}),
\end{align}
where the softmax converts a $4$-dimensional vector to a distribution over four classes, matrix $\mathbf{W}$ and vector $\mathbf{b}$ are trainable parameters, and $\mathbf{s}^{(L)}_{x_m}$ is the output of a stacked of $L$-layer gated convolutional neural
network~\cite{gehring2017convolutional} where the $l$-th layer is:
\[
\mathbf{s}_{x_m}^{(l)} = \mathbf{s}_{x_m}^{(l-1)}+\mathrm{conv}_1(\mathbf{s}_{x_m}^{(l-1)})\otimes\sigma(\mathrm{conv}_2(\mathbf{s}_{x_m}^{(l-1)})),
\]
for $l=1, \cdots, L$, and $\otimes$ denotes element-wise multiplication, $\text{conv}_1$ and $\text{conv}_2$ denote two 
dilate convolutional neural Network~\cite{yu2015multi} with the same convolution kernel size. 
Note that the use of a stacked gated convolutional neural network enables the model to capture farther distances without increasing model parameters, which effectively addresses the issue caused by a large number of sentences in a legal case. 
$\sigma(\cdot)$ denotes a sigmoid gating function controlling which inputs $\text{conv}_1(\mathbf{s}_{x_m}^{(l-1)})$ of the current context are relevant. 
In the first layer, $\mathbf{s}^{(0)}_{x_m}$ is obtained by adding a trainable one-layer multi-layer perceptron on the frozen sentence embedding $x_m$.

Similarly, given a legal case $Y$, the legal characteristics of each sentence embedding $y_n$ in $Y$ can also be identified by classifying $Y$ with the same sentence representations model and neural networks defined above. 
As a result, the rationale characteristic matrix $\mathbf{C}^r=[c_{mn}^r]\in \{0, 1\}^{M\times N}$ can be defined to explicitly indicate whether two sentences have the same predicted legal characteristics:
\begin{equation}\label{eq:class}
    \mathbf{C}^r=\mathbf{M}\otimes (\mathbf{\tilde r}^{X}(\mathbf{\tilde r}^{Y})^{\top}),
\end{equation}
where $\mathbf{M}\in \{0, 1\}^{M\times N}$ is a mask matrix filtering out the padding sentences,
$\mathbf{\tilde r}^{X}\in\{0, 1\}^{M\times 4}$ and $\mathbf{\tilde r}^{Y}\in\{0, 1\}^{N\times 4}$ are the rationale label matrix, whose rows are one-hot representations of $\mathbf{\hat r}^{X}$ and $\mathbf{\hat r}^{X}$.
To incorporate Eq.~\eqref{eq:argmax} into Eq.~\eqref{eq:class} in a differentiable manner, we apply the Straight-Through Gumbel Trick~\cite{bengio2013estimating} to derive $\mathbf{\tilde r}^X$ and $\mathbf{\tilde r}^Y$. 
Accordingly, $c^{r}_{mn}=1$ means that the $m$-th sentence in $X$ and the $n$-th sentence in $Y$ are identified as rationales (i.e., $\hat r_{x_m}\neq 0$ and $\hat r_{y_n}\neq 0$) and they belong to the same rationale type ($\hat r_{x_m}=\hat r_{y_n}$).

\subsection{Generating Candidate Explanations}
\label{sec:generation}
As aforementioned, the optimal transport $\mathbf{A}^*$ indicates pro and con rationale pairs, and thus, can help to generate explanations (i.e., $e$'s) to support matching results (i.e, $z$'s). 
Following the work in~\cite{kumar2020nile}, our IOT-Match exploits the existing pre-trained language model\footnote{We adopt Chinese T5-PEGASUS model: \url{https://github.com/ZhuiyiTechnology/t5-pegasus}. Note that other pre-trained language models are also applicable.} to build three label-specific explanation generators, that is: $f_2 =\{{G}_z\}, z=2, 1, 0$ respectively corresponds to matched, partially matched, and mismatched decisions as shown in Eq.~\eqref{eq:zlabel}. 

The three generators are fine-tuned separately. For example, for $z=0$, the data for fine-tuning ${G}_0$ is selected from the training corpus: $\mathcal{D}_0 = \{(X, Y, \mathbf{\hat{r}}^X, \mathbf{\hat{r}}^Y, e, z=0)\}\subseteq \mathcal{D}$. Given each instance in $\mathcal{D}_0$, it is converted to the input sequence ``$[x_{input};y_{input}]$'' which is a concatenation of two text sequences: $x_{input} =[T_1; x_1; \cdots ;T_m; x_m]$ and $y_{input} =[T_1; y_1; \cdots ; T_n; y_n]$, where $x_i$ ($y_j$) is the sentence of the $i$-th ($j$-th) rationale in $X$ ($Y$), $T_i$ ($T_j$) is the special token indicating the rationale type~\footnote{Six special tokens ``[AI]'', ``[AO]'', ``[YI]'', ``[YO]'',``[ZI]'', and ``[ZO]'' are defined, where A, Y, and Z stand for the key circumstance, constitutive elements of crime, and focus of disputes; I and O stand for pro and con. Non-rationale sentences were discarded.}, and $m$ ($n$) is number of identified rationales.
To fine-tune the parameters in $G_0$, the language modeling loss~\cite{raffel2019exploring} that compares difference between the generated explanation $G_0([x_{input};y_{input}])$ and the human-annotated explanation $e$ is optimized. Similarly, $G_1$ and $G_2$ are fine-tuned based on corresponding subsets $\mathcal{D}_1$ and $\mathcal{D}_2$.


At explanation generation phase, given a tuple $(X, Y, \mathbf{\hat{r}}^X, \mathbf{\hat{r}}^Y)$, we feed the constructed input text sequence $[x_{input};y_{input}]$ to $G_0$, $G_1$, and $G_2$, generating three candidate explanations $\hat e_0, \hat e_1$ and $\hat e_2$.



\subsection{Matching Prediction}
\label{sec:matching}
Instead of considering all of  the sentences in the paired legal cases, IOT-Match
learns the $f_3$ to conducts matching only based on the extracted rationales as well as the generated candidate explanations. 
This strategy makes the extracted rationales and generated explanations faithful to the matching predictions and avoids the negative impact of noise sentences on the matching results. 
Formally, given a paired legal case $(X,Y)$, our IOT-Match would identify their relation by calculating a probabilistic distribution over the three classes shown in Eq.~\eqref{eq:zlabel}:
\begin{equation}
\label{eq:matching}
\begin{aligned}
\{P(z=k|(X,Y)\}_{k=0}^2=\mathrm{softmax}(\mathbf{W}[{\hat{z}}_0;{\hat{z}}_1;{\hat{z}}_2]+\mathbf{b}),
\end{aligned}
\end{equation}
where $[;]$ concatenates vectors, and $\mathbf{W}$ and $\mathbf{b}$ are trainable parameters. 
As for $\mathbf{\hat{z}}_i$ $(i\in\{0,1,2\})$, following the practice in~\cite{kumar2020nile,zhao2021lirex}, the matching scores are computed based on the extracted rationales and the corresponding candidate explanations:
\[
 {\hat z}_i=\mathrm{MLP}([\mathbf{s}_{r^X};\mathbf{s}_{r^Y}; \mathbf{s}_{\hat e_i}]),
\]
where $\mathbf{s}_{r^X}, \mathbf{s}_{r^Y}, \mathbf{s}_{\hat e_i}$ respectively denote the embeddings of rationales of $X$, $Y$, and the candidate explanation $\hat e_i$ which are obtained by tuning the BERT model\footnote{\url{https://github.com/thunlp/OpenCLaP}}; 
$\text{MLP}$ denotes a two-layer perceptron with sigmoid activation functions. 
Accordingly, our IOT-Match makes the final matching decision for a paired legal case $(X,Y)$ as
\begin{equation}
    \hat{z}=\mathop{\arg\sideset{}{_{k\in\{0,1,2\}}}\max}P(z=k|(X,Y)),
\end{equation}
and outputs the explanation corresponding to the highest matching score at the same time.

\subsection{Model Training}
\label{sec:training}
IOT-Match has parameters to determine during the training, including those in the pro and con rationales extraction ($f_1$), those in the candidate explanations generation ($f_2$), and those in the matching ($f_3$). 
These models parameters respectively denoted as $\theta_{f_1}, \theta_{f_2}, \theta_{f_3}$ are trained sequentially, and the output of the $f_1$ is used as the input of the $f_2$, and the output of the $f_1$ and $f_2$ are used as the input of the $f_3$. The training process of IOT-Match is illustrated in Algorithm~\ref{alg:training}.
\begin{algorithm}[!t] 
    \caption{Training Process of IOT-Match.}
    \label{alg:training} 
    \begin{algorithmic}[1]
    \REQUIRE Training set $\mathcal{D}=\{(X_i, Y_i, \mathbf{r}^X_i, \mathbf{r}^Y_i, \hat{\mathbf{A}}_i,z_i,e_i)\}_{i=1}^N$; mini-batch sizes $n_1, n_2, n_3$;  trade-off coefficients $\epsilon, \gamma_1, \gamma_2, \gamma_3$; entropic regularizer coefficient $\gamma$; learning rates $\eta_1, \eta_2, \eta_3$.
    \STATE $\rhd$ IOT-based Rationale Extraction
    \REPEAT 
    \STATE Sample a mini-batch $\{(X_i, Y_i, \mathbf{r}^X_i, \mathbf{r}^Y_i, \hat{\mathbf{A}}_i)\}^{n_1}_{i=1}$ from $\mathcal{D}$
    \STATE Predict rationale labels $\hat{\mathbf{r}}^X_i, \hat{\mathbf{r}}^Y_i$ for $(X_i,Y_i)$ 
    \COMMENT{Eq.~\eqref{eq:ClassDistr}}
    \STATE Calculate $\mathcal{L_R}$ \COMMENT{Eq.~\eqref{eq:lr}}
    \STATE Construct $\mathbf{C}^s, \mathbf{C}^r, \mathbf{C}$ and optimize $\mathbf{A}^*$ by Sinkhorn scaling
    \STATE Calculate $\mathcal{L_A}$, with $\mathbf{A}^*=\mathbf{A}^*(\mathbf{C})$ 
    \COMMENT{Eq.~\eqref{eq:semi}}
 \STATE
 $\mathcal{L}_{f_1}=\sum_1^{n_1}\mathcal{L_R} + \gamma_1\mathcal{L_A}$
    \COMMENT{Eq.~\eqref{eq:f1}}
    \STATE $\theta_{f_1}\leftarrow \theta_{f_1} - \eta_1  \bigtriangledown_{\theta_{f_1}} \mathcal{L}_{f_1}$
    \UNTIL{convergence} 
    \RETURN $\theta_{f_1},\{\hat{\mathbf{r}}^X, \hat{\mathbf{r}}^Y\}$
     \STATE $\rhd$ Generating Candidate Explanations
     \REPEAT
     \STATE Sample a mini-batch $\{(X_i, Y_i, \hat{\mathbf{r}}^X_i, \hat{\mathbf{r}}^Y_i,  e_i)\}^{n_2}_{i=1}$ from $\mathcal{D}$
     \STATE Calculate $\mathcal{L}_{f_2}$\COMMENT{Eq.~\eqref{eq:f2}}
     \STATE Fine-tune three label-specific pre-trained language models.
    \UNTIL{convergence} 
    \RETURN $\theta_{f_2},\{\hat{e}_0, \hat{e}_1, \hat{e}_2\}$
    \STATE $\rhd$ Matching Prediction
    \REPEAT
    \STATE Sample a mini-batch $\{(X_i, Y_i, \hat{\mathbf{r}}^X_i, \hat{\mathbf{r}}^Y_i, z_i, e_i, \hat{e}_i)\}^{n_3}_{i=1}$ from $\mathcal{D}$
    \STATE
    Predict matching label $\hat{z}$ using predicted rationales candidate explanations $\hat{e}_0, \hat{e}_1, \hat{e}_2$
    \COMMENT{Eq.~\eqref{eq:matching}}
    \STATE Calculate $\mathcal{L_M}$
    \COMMENT{Eq.~\eqref{eq:lm}}, $\mathcal{L_E}$
    \COMMENT{Eq.~\eqref{eq:le}}, and $\mathcal{L_C}$ \COMMENT{Eq.~\eqref{eq:lc}}.
    \STATE $\mathcal{L}_{f_3}=\sum_1^{n_3}\mathcal{L_M} + \gamma_3(\mathcal{L_E} + \mathcal{L_C})$
    \COMMENT{Eq.~\eqref{eq:f3}}
    \STATE $\theta_{f_3}\leftarrow \theta_{f_3} - \eta_3  \bigtriangledown_{\theta_{f_3}} \mathcal{L}_{f_3}$
    \UNTIL{convergence} 
    \RETURN $\theta_{f_3}$
    \end{algorithmic}
    \end{algorithm}
Specifically, in the $f_1$, the learning objective is defined to measure the loss of the pro and con rationales extraction:
\begin{equation}
\label{eq:f1}
    \mathcal{L}_{f_1}=\sideset{}{_{(X,Y, \mathbf{r}^X, \mathbf{r}^Y,\mathbf{\hat{A}})\in\mathcal{D}}}\sum\mathcal{L_R} + \gamma_1\mathcal{L_A},
\end{equation}
where, for each legal case pair in the dataset, the loss function consists of two parts: the rationale identification loss $\mathcal{L_R}$ and the affinity matrix loss $\mathcal{L_A}$. 
The $\gamma_1>0$ is a hyper-parameter controlling their weights. 
The rationale identification loss $\mathcal{L_R}$ is defined as the cross-entropy loss between the ground-truth rationale labels of each sentence and the corresponding predictions:
\begin{equation}
\label{eq:lr}
\begin{aligned}
 \mathcal{L_R}=&-\sideset{}{_{k=0}^{3}}\sum\Bigl(\sideset{}{_{m=1}^M}\sum\delta(r_{x_m},k)\log(P(\hat r_{x_m}=k|x_m))\\
 &+\sideset{}{_{n=1}^N}\sum\delta(r_{y_n},k)\log(P(\hat r_{y_n}=k|y_n))\Bigr),
\end{aligned}
\end{equation}
where $\delta(r,k) =1$ if $r=k$ else 0.
The loss $\mathcal{L_A}$ is based on the IOT problem in Eq.~\eqref{eq:IOT4Legal}:
\begin{equation}
\begin{aligned}
\label{eq:semi}
    \mathcal{L_A}=\mathrm{KL}(\mathbf{\hat{A}}||\mathbf{A}^*(\mathbf{C})) + \gamma_2\sideset{}{_{m=1}^M}\sum\sideset{}{_{n=1}^N}\sum\delta(\hat{r}_{x_m},\hat{r}_{y_n})c_{mn},
\end{aligned}
\end{equation}  
where the first term corresponds to the IOT problem that optimize the affinity matrix and the associated optimal transport to fit a small number of alignment labels (i.e., $\mathbf{\hat{A}}$). 
The second term is an unsupervised loss based on the predicted rationale labels, which explicitly regularizes the affinity matrix $\mathbf{C}$ to minimize the discrepancy between identical rationales and maximize the that between different rationales.
Here, $\delta(\hat{r}_{x_m},\hat{r}_{y_N})=1$ if $\hat{r}_{x_m}=\hat{r}_{y_N}\neq 0$ else 0, $\gamma_2$ is a coefficient to balance the supervised loss and the unsupervised loss.

In the $f_2$, its learning objective $\mathcal{L}_{f_2}$ is identical to that used in the fine-tuning phase of the pre-trained language models~\cite{raffel2019exploring,zhang2020pegasus,kumar2020nile}:
\begin{equation}
    \label{eq:f2}
    \mathcal{L}_{f_2}=-\sideset{}{_{(X, Y, \hat{\mathbf{r}}^X, \hat{\mathbf{r}}^Y, e)\in\mathcal{D}}}\sum\sideset{}{_{l=1}^L}\sum\log (P(s_l|\mathbf{s}_{1:l-1})),
\end{equation}
where $\mathbf{s}$ stands for a sample sequence which contains $L$ tokens, $s_l$ denotes for the $l$-th token of $\mathbf{s}$, and $\mathbf{s}_{1:l-1}$ denotes the prefix of $s_l$.

In the $f_3$, the loss function consists of three parts:
\begin{equation}
\label{eq:f3}
    \mathcal{L}_{f_3}=\sideset{}{_{(X, Y, \hat{\mathbf{r}}^X, \hat{\mathbf{r}}^Y, z, e,\hat{e})\in\mathcal{D}}}\sum\mathcal{L_M} + \gamma_3(\mathcal{L_E} + \mathcal{L_C}),
\end{equation}
where $\gamma_3>0$ is a coefficient to balance $\mathcal{L_M}$, $\mathcal{L_E}$ and $\mathcal{L_C}$.
$\mathcal{L_C}$ is 
the cross-entropy loss between the ground-truth matching label $z$ and the matching score of rationales and candidate explanations:
\begin{equation}
    \label{eq:lm}
    \mathcal{L_M}=-\sideset{}{_{k=0}^{2}}\sum \delta(z,k)\log\left(P(\hat z_k=k|(X,Y))\right),
\end{equation}
where $\delta(z,k) =1$ if $z=k$ else 0.

We also design two auxiliary tasks for learning a better representation for rationales and explanations. 
To ensure that the human-annotated explanation $e$ accurately reflects the matching relation between rationales, the similarity between $[\mathbf{s}_{r^X}, \mathbf{s}_{r^Y}]$ and $\mathbf{s}_e$ should be larger than that between $[\mathbf{s}_{r^X}, \mathbf{s}_{r^Y}]$ and the generated explanation $\mathbf{s}_{{\hat e}_k}$.
Therefore, the first task is designed as:
\begin{equation}
\label{eq:le}
\begin{aligned}
    \mathcal{L_E}=\sideset{}{_{k=0}^2}\sum\max\Big(0,&\cos(\mathrm{MLP}[\mathbf{s}_{r^X};\mathbf{s}_{r^Y}],\mathbf{s}_{\hat e_k})\\
    &-\cos(\mathrm{MLP}[\mathbf{s}_{r^X}; \mathbf{s}_{r^Y}],\mathbf{s}_{ e})\Big),
\end{aligned}
\end{equation}
where MLP denotes a one-layer multi-layer perceptron.
Moreover, inspired by the success of contrastive learning~\cite{chen2020simple,gao2021simcse,liu2021simcls} and the observations in~\cite{kumar2020nile} that explanations with the same label tend to have the same form, and the form of explanations may be the noise for matching, the second auxiliary task is designed to avoid the classifier only using the form of explanations to infer the matching prediction. 
Specifically, the candidate explanations in current data are regarded as positive samples $\mathbf{s}_{{\hat e}_k}$, and the explanations with the same label in the mini-batch are regarded as negative samples $\mathbf{s}_{\hat e^l_k-}$. 
Then, the cosine similarity between rationales and positive/negative explanations are calculated and compared:
\begin{equation}
\label{eq:lc}
    \begin{aligned}
    \mathcal{L_C}=\sideset{}{_{k=0}^2}\sum\sideset{}{_l}\sum\max\Big(0,&\cos(\mathrm{MLP}[\mathbf{s}_{r^X};\mathbf{s}_{r^Y}],\mathbf{s}_{\hat{e}^l_k-})\\
    &-\cos(\mathrm{MLP}[\mathbf{s}_{r^X}; \mathbf{s}_{r^Y}],\mathbf{s}_{\hat e_k})\Big),
\end{aligned}
\end{equation}
where $l$ is the number of negative samples. 

\section{ELAM and eCAIL: New Datasets for Explainable Legal Case Matching}
\label{sec:elam}

To verify the effectiveness of our IOT-Match method, we construct a new Explainable Legal cAse Matching (ELAM) dataset that provides not only the ground-truth matching label for each legal case pair, but also manually-labeled rationales, their alignments, and natural language-based explanations of the matching decision. 

To construct the ELAM dataset, we collect 8955 legal cases on ``the obstruction of social management order crime'' from Faxin\footnote{\url{https://www.faxin.cn}}. 
Each case is already associated with several tags that provide some basic information such as application of the law, court name, judge year, trial or retrial, etc.
During the pre-processing, we randomly sample 1250 cases as queries and construct a candidate pool for each query case. 
The cases in the candidate pool are retrieved according to their numbers of overlapped tags compared to the corresponding query case. 
To ensure the usability of the dataset, we remove those query cases (and their candidate pools) when the candidate pool retrieved less than 10 cases. 
In total, the final dataset contains 5000 legal case pairs. 
Finally, the basic information (e.g., the tags) is removed and some identity information is replaced with placeholders for privacy protection.

During the human annotation, each legal expert is provided a set of randomly selected legal case pairs. 
For each pair, a legal expert was asked to first annotate the rationale label for each sentence. Following the practices in~\cite{ma2021lecard}, the rationale labels are 4-level: the key circumstances, the key
constitutive elements of crime, the focus of disputes, and not a rationale.
Then, the pro and con rationales (the alignment of rationales) were marked. 
Finally, the 3-level matching label was annotated.
In addition, the legal experts were required to give a free-form natural language explanation for their matching decision based on the annotated rationales. 
\begin{table}[t]
    \centering
    \caption{Statistics of ELAM and eCAIL. 
    Types of sentences include rationales and the others.
    \#Rationale denotes the average number of rationales for each type per case; 
    Prop. of pro and con denotes the average proportion of pro and con rationales per case.
    }
    \begin{tabular}{l|c|c}
        \toprule
        &ELAM&eCAIL\\
        \midrule
         Train/ Valid/ Test&4000/ 500/ 500 & 6000/ 750/ 750\\
         Types of sentences& 4&2\\
         \#Rationale &4.80/6.00/ 1.30& 12.00\\
         Prop. of pro and con&2.61:7.39& 3.52:6.48
\\
         \#Sentence per case&16.29&124.10\\
         Length of explanation&176.34&165.39\\
         \bottomrule
    \end{tabular}
    
    \label{tab:data}
\end{table}

Besides ELAM, we also extended the CAIL (Challenge of AI in Law) 2021 dataset to adapt to the explainable legal case matching task.
This dataset is created for the NLP competition in the law domain.\footnote{We use the Fact Prediction Track data available at:~\url{http://cail.cipsc.org.cn/}}
Each legal case in the original CAIL corpus is associated with several tags about the issue of private lending. 
In our extended CAIL (eCAIL), these tagged sentences are regarded as rationales, and others are regarded as unrelated sentences. 
The pro and con rationales correspond to the tagged sentences with identical labels and those with different labels, respectively.
The same pre-processing as that of ELAM is conducted to construct the candidate legal case pairs. 
Because the tags in eCAIL data are the constitutive elements of crime for rationale sentences, we automatically assign a matching label for a pair of cases according to their tag-overlapping (overlapping $>10$ means matching, overlapping $<1$ means mismatching, else means partially matching). 
Similarly, as for the natural language-based explanation of the matching label, we concatenate all of the tags in a paired legal case. 
Some basic statistics of ELAM and eCAIL are listed in Table~\ref{tab:data}.

\begin{table*}[t]
    \small
\caption{Experimental results on ELAM and eCAIL test sets. 
    $^\dagger$ indicates the statistically significant difference between the performance of all baseline models and that of IOT-Match ($p\textrm{-value}< 0.05$).
    }
    \label{tab:Exp:matching}
\centering
    \begin{tabular}{l|l|rrrr|rrrr}    
        \toprule
        &&\multicolumn{4}{c|}{ELAM}
        &\multicolumn{4}{c}{eCAIL}
        \\
        \textbf{Model types}&
         \textbf{Models}  & \textbf{Acc. (\%)} & \textbf{P. (\%)}& \textbf{R. (\%)} & \textbf{F1 (\%)}& \textbf{Acc. (\%)} & \textbf{P. (\%)}& \textbf{R. (\%)} & \textbf{F1 (\%)}\\
        \midrule
        &Sentence-BERT~\cite{reimers2019sentence}&68.83	&69.83& 66.88& 67.20&
        71.33& 70.83& 71.21&70.98\\
        ~&Lawformer~\cite{xiao2021lawformer}&
        69.91& 72.26&	68.34&	69.18&
        70.67&	70.20&	70.55&	69.91
        \\
        Legal case&BERT-PLI~\cite{shao2020bert}&
        71.21&	71.22&	71.23&	70.88&
        70.66&	70.05&	70.54&	70.18\\
        matching&Thematic Similarity (avg)~\cite{bhattacharya2020methods}&
        70.99	& 71.28& 68.97& 69.12&
        71.47& 70.88& 71.34& 71.00
        \\
        ~&Thematic Similarity (max)~\cite{bhattacharya2020methods}&
        71.86& 71.50&70.07& 70.26&
        68.53& 67.25& 68.38	& 67.57\\
        \midrule
        ~&NILE (Agg)~\cite{kumar2020nile}&65.87& 65.22&64.89&	65.05&
        71.60&	71.44&	71.02&	70.91\\
        Short text matching&NILE (App)~\cite{kumar2020nile}&
        68.90 &68.90& 66.87&67.32&
        72.53&	71.97&	71.93&	71.95
        \\
        with explanations&NILE (Ind)~\cite{kumar2020nile}&69.76&	68.30&	68.82&	68.46
        &73.33&	73.43&	72.84&	73.05\\
        ~&LIREx~\cite{zhao2021lirex}&
        68.18&	68.22&	67.34&	67.66&
        70.53&	69.68&	70.40&	69.94
        \\
        \midrule
        Ours&IOT-Match&\textbf{73.87}$^\dagger$&	\textbf{73.02}$^\dagger$&	\textbf{72.41}$^\dagger$&	\textbf{72.55}$^\dagger$& \textbf{82.00}$^\dagger$&	\textbf{82.10}$^\dagger$&	\textbf{81.92}$^\dagger$&	\textbf{81.90}$^\dagger$\\
        \bottomrule
    \end{tabular}
\end{table*}

\section{Experiments}
In this section, we conduct experiments to answer the following research questions: 
\textbf{RQ1:} Can IOT-Match outperform state-of-the-art methods on legal case matching and text matching with explanations in terms of matching accuracy? 
\textbf{RQ2:} How good are the explanations produced by IOT-Match, including the extracted rationales and the generated natural language?
\textbf{RQ3:} Can IOT-Match efficiently make use of limited rationale alignment labels? 

The source code, ELAM and eCAIL datasets, and all experiments have been shared at:~\url{https://github.com/ruc-wjyu/IOT-Match}.

\subsection{Experimental Settings}
\subsubsection{Baselines and Evaluation Metrics}
To the best of our knowledge, there exist few models that are exactly designed for explainable legal case matching. 
In the experiments, two types of text matching models are selected as baselines. 
The first type includes state-of-the-art legal case matching models without explanations: 
\\1) \textbf{Sentence-BERT}~\cite{reimers2019sentence} uses BERT pre-trained on the legal case corpus\footnote{\url{https://github.com/thunlp/OpenCLaP}} to encode two cases and uses a MLP to conduct matching. \\
2) \textbf{Lawformer}~\cite{xiao2021lawformer} leverages a Longformer-based~\cite{beltagy2020longformer} pre-trained language model for Chinese legal long documents understanding. \\
3) \textbf{BERT-PLI}~\cite{shao2020bert} uses BERT to capture paragraph-level semantic relations and then aggregates them with RNN and attention.\\
4) \textbf{Thematic Similarity}~\cite{bhattacharya2020methods} segments two legal cases into paragraphs and computes the paragraph-level similarities. Maximum or average similarities are used for the overall matching prediction.

The second type of baselines includes the following matching models designed for short text matching with explanations: \\
1) \textbf{NILE}~\cite{kumar2020nile} adopts GPT2 to generate label-specific explanations for paired sentences, which has three variants that leverage different information to output matching score: \textbf{NILE (Ind)} only uses the generated explanation; \textbf{NILE (App)} uses the concatenation of input paired sentences and the generated explanation; and \mbox{\textbf{NILE (Agg)}} compares all the generated label-specific explanations.\\
2) \textbf{LIREx}~\cite{zhao2021lirex} uses an attention mechanism to generate rationale-enabled explanations, which also involves selected explanations to conduct the sentence matching.

Note that both ELAM and eCAIL are in Chinese and do not have precedent information, we do not choose the precedent citation network-based methods~\cite{bhattacharya2020methods,bhattacharya2020hier} as the baselines. 
Additionally, since NILE and LIREx can generate natural language explanations for matching, we compared IOT-Match with them in terms of explanation generation using  identical pre-trained language models.

To evaluate the performance of rationale extraction, we also compare IOT-Match with the following state-of-the-art rationale extraction models designed for paired documents:\\ 1) \textbf{MT-H-LSTM}~\cite{cheng2020argument} uses two bi-LSTMs to obtain sentence embeddings and predict the aligned sentences from document pairs.\\
2) \textbf{MLMC}~\cite{cheng2021argument} formulates the associative sentence extraction for paired documents as a problem of table filling, in which a matrix is constructed to show whether the sentences are related or not.\\
3) \textbf{DecAtt}~\cite{parikh2016decomposable} adopts attention to indicate the alignments between cross-case sentences. 
To make fair comparisons, the sentence encoder in DecAtt is set to be identical to that of in IOT-Match.

Different metrics are adopted to evaluate the different modules of IOT-Match. 
As for rationales extraction and matching prediction, Accuracy, Precision, Recall, and F1 are used. 
As for natural language explanation generation, the ROUGE score is used because the task is formulated as the Seq2Seq text generation. 

\subsubsection{Hyper-parameter settings}
All of the hyper-parameters in IOT-Match are tuned using grid search on the validation set with Adam~\cite{kingma2014adam}. 
In the rationale extraction, the learning rate $\eta_1$ is tuned between $\{1e-4, 1e-3\}$; 
the batch size $n_1$ is tuned among $\{32, 64, 128\}$; 
$\gamma_1$ is tuned between $[1, 10]$ and $\gamma_2$ is tuned between $[0.1, 1.0]$;
the alignment threshold $\tau$ for ELAM and eCAIL are tuned between $[1e-3,1e-2]$ and $[1e-3,5e-3]$, respectively; and the entropic regularizer $\gamma$ is tuned among $[0.1, 1.0]$;
the affinity matrix coefficient $\epsilon$ is tuned among $\{0, -10, -50, -100, -200\}$. 
In the natural explanation generation, the hyper-parameters are set according to those reported in~\cite{zhuiyit5pegasus}:
the learning rate $\eta_2$ is set as $2e-5$; the batch size $n_2$ is set as $2$;
In the matching, the learning rate $\eta_3$ is tuned between $\{2e-5,2e-4\}$; the batch size $n_3$ is tuned between $\{4, 8\}$, and $\gamma_3$ is tuned among $\{1,10,20\}$.

\subsection{Matching Accuracy (RQ1)}
We first study the matching performance of our proposed IOT-Match.
Table~\ref{tab:Exp:matching} presents the matching performances of IOT-Match and the baselines in terms of four evaluation metrics on ELAM and eCAIL. 
All the methods are trained ten times and the averaged results are reported. 
Based on the results, we summarize our observations as follows:
(1) IOT-Match consistently and significantly outperforms all of the baselines on two datasets in terms of all metrics, indicating the effectiveness of IOT-Match in enhancing the matching accuracy. 
Note that the legal cases in eCAIL are extremely lengthy (on average over 100 sentences per legal case). 
IOT-Match achieves over 10\% improvements in terms of all four metrics on eCAIL, further verifying its effectiveness in the matching of long-form legal cases.
(2) Compared to short text matching with explanation models which involve all sentences in a paired legal case during the matching, IOT-Match enjoys the advantages from the extracted rationales and achieves consistent improvements on two datasets. 
The result indicates that the rationale extraction module in IOT-Match accurately identified the rationales and filtered out the noise sentences from legal cases.
(3) Compared to existing legal case matching models that cannot provide matching explanations, IOT-Match also achieves consistent improvements on both datasets.
The results indicate that the natural language explanations generated by IOT-Match are helpful for legal case matching.

\subsection{\mbox{Quality of Rationales and Explanations (RQ2)}}
The major superiority of IOT-Match compared to existing legal case matching models is that IOT-Match is able to extract rationales and generate explanations for the matching prediction. 
In this subsection, we conduct experiments to assess the quality of the extracted rationales and the generated natural language explanation by IOT-Match. 
Following~\citep{deyoung2020eraser}, we adopt plausibility and faithfulness as the metrics. 
Plausibility measures how well the explanation aligns with human annotations, and faithfulness measures the degree to which the explanation influences the corresponding predictions. 


\begin{table}[t]
    \centering
    \caption{Plausibility of extracted rationales on ELAM and eCAIL test sets in terms of extraction accuracy.
    }    \label{tab:exp:rationale_plausibility}
    \begin{tabular}{l|c|c}    
        \toprule
        &{ELAM}&{eCAIL}\\
        \textbf{Models} &\textbf{Acc. (\%)}&\textbf{Acc. (\%)}
        \\
        \midrule
        MT-H-LSTM~\cite{cheng2020argument}&68.91&95.18\\
        MLMC~\cite{cheng2021argument}&68.37&95.30\\
        DecAtt~\cite{parikh2016decomposable}&83.09&94.33\\
        OT~\cite{peyre2019computational}&83.09&90.97\\
        IOT-Match&\textbf{86.82}&\textbf{96.26}\\
        \bottomrule
    \end{tabular}
\end{table}

\subsubsection{Quality of the extracted rationales}
In terms of \textbf{plausibility}, we compare the rationales extracted by IOT-Match and baseline models with human annotations on ELAM and eCAIL. 
As shown in Table~\ref{tab:exp:rationale_plausibility}, the rationales extracted by IOT-Match are more consistent with human annotations, especially on the ELAM dataset where the rationales are more diverse (three types of rationales). 
According to the results illustrated in the figure, IOT-Match outperforms MLMC~\cite{cheng2021argument}, MT-H-LSTM~\cite{cheng2020argument} and DecAtt~\citep{parikh2016decomposable} by about 24.0\%, 25.0\%, and 4.0\%, respectively on ELAM.
We also compare the original IOT-Match with a modified one with IOT ablated but forward OT kept, denoted as OT in Table~\ref{tab:exp:rationale_plausibility}. 
Table~\ref{tab:exp:rationale_plausibility} shows that the extraction accuracy drops if we remove IOT from IOT-Match. 
The results indicate the effectiveness of IOT in learning the adaptive cross-case sentence affinity and predicting the rationale alignment.

\begin{table}[t]
    \small
    \centering
    \caption{Faithfulness of the extracted rationales and the generated explanation on ELAM and eCAIL test sets. The column ``Input'' denotes IOT-Match with different inputs.
    }    \label{tab:exp:rationale_faithful}
    \begin{tabular}{@{ }l@{ }|@{ }r@{ }r@{ }r@{ }r@{ }|@{ }r@{ }r@{ }r@{ }r}    
        \toprule
        &\multicolumn{4}{c@{ }|@{ }}{ELAM}
        &\multicolumn{4}{c}{CAIL}
        \\
         \textbf{Input}  & \textbf{Acc. (\%)} & \textbf{P. (\%)}& \textbf{R. (\%)} & \textbf{F1 (\%)}& \textbf{Acc. (\%)} & \textbf{P. (\%)}& \textbf{R. (\%)} & \textbf{F1 (\%)}\\
        \midrule
        $a\backslash r$&65.01&	64.24& 63.29&63.35&72.40	 &71.78&72.30 &71.87\\
         $a$&68.83	&69.83& 66.88& 67.20&
        71.33& 70.83& 71.21&70.98\\
         $r$&70.35&70.06  &68.94&69.22&{72.67}& {72.29}	 &{72.53}&{72.08}\\
        \midrule
         $e$&
        {71.27}&{70.47}&{70.71}&{70.58}&68.47&68.52&65.67&	66.05\\	
         $a\backslash r + e$&69.98&	70.75&	68.79&	69.51&79.20&	79.63&	69.45&	79.14\\
         $a + e$&73.65&	\textbf{73.29}&	\textbf{73.29}&	\textbf{73.26}&71.33&	70.83&	71.21&	70.98\\
         $r + e$&\textbf{73.87}&	73.02&72.41&72.55&\textbf{82.00}&	\textbf{82.10}&	\textbf{81.92}&	\textbf{81.90}\\
        \bottomrule
    \end{tabular}
\end{table}
\begin{table*}[t]
    \small
    \caption{Plausibility of generated explanations on ELAM and eCAIL test sets in terms of ROUGE scores.
    }
    \label{tab:Exp:rouge1}
    \centering
    \begin{tabular}{l|c@{ }c@{ }c|c@{ }c@{ }c|c@{ }c@{ }c|c@{ }c@{ }c|c@{ }c@{ }c|c@{ }c@{ }c}    
        \toprule
        &\multicolumn{9}{c|}{ELAM}&\multicolumn{9}{c}{eCAIL}\\
        &\multicolumn{3}{c|}{\textbf{ROUGE-1 (\%)}}
        &\multicolumn{3}{c|}{\textbf{ROUGE-2 (\%)}}
        &\multicolumn{3}{c|}{\textbf{ROUGE- L (\%)}}
        &\multicolumn{3}{c|}{\textbf{ROUGE-1 (\%)}}
        &\multicolumn{3}{c|}{\textbf{ROUGE-2 (\%)}}
        &\multicolumn{3}{c}{\textbf{ROUGE- L (\%)}}
        \\
       \textbf{Models}  & \textbf{$z=2$} & \textbf{$z=1$}& \textbf{$z=0$} & \textbf{$z=2$} & \textbf{$z=1$}& \textbf{$z=0$} & \textbf{$z=2$} & \textbf{$z=1$}& \textbf{$z=0$} & \textbf{$z=2$} & \textbf{$z=1$}& \textbf{$z=0$} & \textbf{$z=2$} & \textbf{$z=1$}& \textbf{$z=0$} & \textbf{$z=2$} & \textbf{$z=1$}& \textbf{$z=0$}\\
        \midrule
        NILE ~\cite{kumar2020nile}&73.40&70.96&69.93&58.47&55.57&	56.08&69.87&65.70&	66.84&73.40&70.96&69.93&58.47&55.57&	56.08&69.87&65.70&	66.84\\
        LIREx~\cite{zhao2021lirex}&74.15&71.61&70.97&59.78&	56.36&56.88&70.89&66.22&	67.41&80.35&	74.36&	74.46&	72.00&	67.35&	63.58&	76.84&	74.28&	66.98
        \\
        IOT-Match  &
        \textbf{75.55}&\textbf{73.64}&	\textbf{75.18}&	\textbf{60.97}&	\textbf{58.25}&	\textbf{61.84}&	\textbf{72.79}&	\textbf{68.85}&	\textbf{72.54}&
        \textbf{83.54}&	\textbf{76.59}&	\textbf{91.21}&	\textbf{76.48}&	\textbf{69.46}&	\textbf{87.03}&	\textbf{80.37}&	\textbf{76.16}&	\textbf{86.91}\\
        
        \bottomrule
    \end{tabular}
\end{table*}
\begin{table}[t]
    \centering
    \caption{Human evaluations of the explanation quality over 50 randomly sampled data from ELAM and eCAIL by two annotators with the inter-rater agreement of 0.95. 
    }    
    \label{tab:exp:human_evalutation}
    \begin{tabular}{l|c|c|c}    
        \toprule
        &NILE~\cite{kumar2020nile}&LIREx~\cite{zhao2021lirex}&IOT-Match\\
        \midrule
        ELAM&35&41&\textbf{46}\\
        \midrule
        eCAIL&36&38&\textbf{44}\\
        \bottomrule
    \end{tabular}
\end{table}
In terms of \textbf{faithfulness}, we conduct experiments to measure the degree to which the extracted rationales influence the final matching. 
Specifically, we test the matching performance of IOT-Match with explanations and IOT-Match without explanations respectively under three conditions: using all sentences as the input (respectively denoted as ``IOT-Match ($a+e$)'' and ``IOT-Match ($a$)''), using rationale extracted by IOT-Match  as the input (respectively denoted as ``IOT-Match ($r+e$)'' and ``IOT-Match ($r$)''), and using sentences except those extracted by IOT-Match as the input (respectively denoted as ``IOT-Match ($a\backslash r+e$)'' and ``IOT-Match ($a\backslash r$)''). 
From the results reported in  Table~\ref{tab:exp:rationale_faithful}, we find that the rationales extracted by IOT-Match play a critical role in legal case matching. 
Specifically, if the extracted rationales are removed from a model's input (IOT-Match($a\backslash r+e$) or IOT-Match($a\backslash r$)), the matching accuracy of the model drops dramatically. 
In addition, in eCAIL where the legal cases are extremely lengthy,
if all sentences are used as a model's input (IOT-Match($a+e$) or IOT-Match($a$)), the model's accuracy still drops to some extent because of the noise from other sentences. 
On ELAM, the performance of using rationales as the only input is competitive with that of using all sentences. 
This result indicates the rationales extracted by IOT-Match already provide sufficient legal semantics for case matching.
Based on the above analysis, we conclude that IOT-Match is capable of accurately extracting faithful rationales for legal case matching.

\subsubsection{Quality of the Generated Explanation}
        
In terms of \textbf{plausibility},
we compare the natural language explanation generated by IOT-Match to those generated by  NILE~\cite{kumar2020nile} and LIREx~\cite{zhao2021lirex}. 
Since both ELAM and eCAIL have human-annotated explanations for the matching labels, the popular metrics in machine translation such as ROUGE-1, ROUGE-2, and ROUGE-L are used to evaluate the plausibility.
As shown in Table~\ref{tab:Exp:rouge1}, the natural language explanations generated by IOT-Match are more consistent with human annotations than those generated by NILE and LIREx, especially on the eCAIL where IOT-Match outperformed NILE~\cite{kumar2020nile} and LIREx~\cite{zhao2021lirex} at least by 7.9\% and 2.5\% across all metrics, respectively. 
Moreover, we also conduct human evaluations to test the quality of the generated explanations.
Following~\cite{zhao2021lirex}, we randomly sampled 50 examples respectively from ELAM and eCAIL, and ask two annotators to answer the questions that whether the generated explanation and the label explanation convey the same meaning.
Each annotator was provided with the context (legal cases, rationales, explanations), and asked to label them as 1 if they agree to the question, or 0 otherwise. As shown in Table~\ref{tab:exp:human_evalutation}, IOT-Match obtains a high relevance score between the generated explanations and label explanations. The result verifies the effectiveness of IOT-Match in generating plausible explanations.

In terms of \textbf{faithfulness}, we conduct experiments to measure the degree to which the generated explanation influences the final matching. 
Specifically, we compare the performance among IOT-Match using the rationales only (IOT-Match($r$)), using the explanation only (IOT-Match($e$)), and using rationales and explanations (IOT-Match($r+e$)). 
From the results reported in Table~\ref{tab:exp:rationale_faithful}, we find: (1) on both ELAM and eCAIL, IOT-Match ($r+e$) performs the best, indicating that the natural language explanations generated by IOT-Match contributed to the matching prediction; 
(2) IOT-Match ($e$) performs better than IOT-Match ($r$) on ELAM, verifying the faithfulness of the generated explanation. 
The result also indicates that the explanations on ELAM are more sufficient for the matching prediction than the extracted rationales.
(3) IOT-Match ($e$) performs worse than IOT-Match ($r$) on eCAIL. 
We analyze the reasons and find that the labeled explanations on eCAIL are the concatenations of the rationale sentences. 
Such labeled explanations are not coherent enough and may harm the generated explanations.
\begin{figure}
    \centering
    \includegraphics[width=0.9\linewidth]{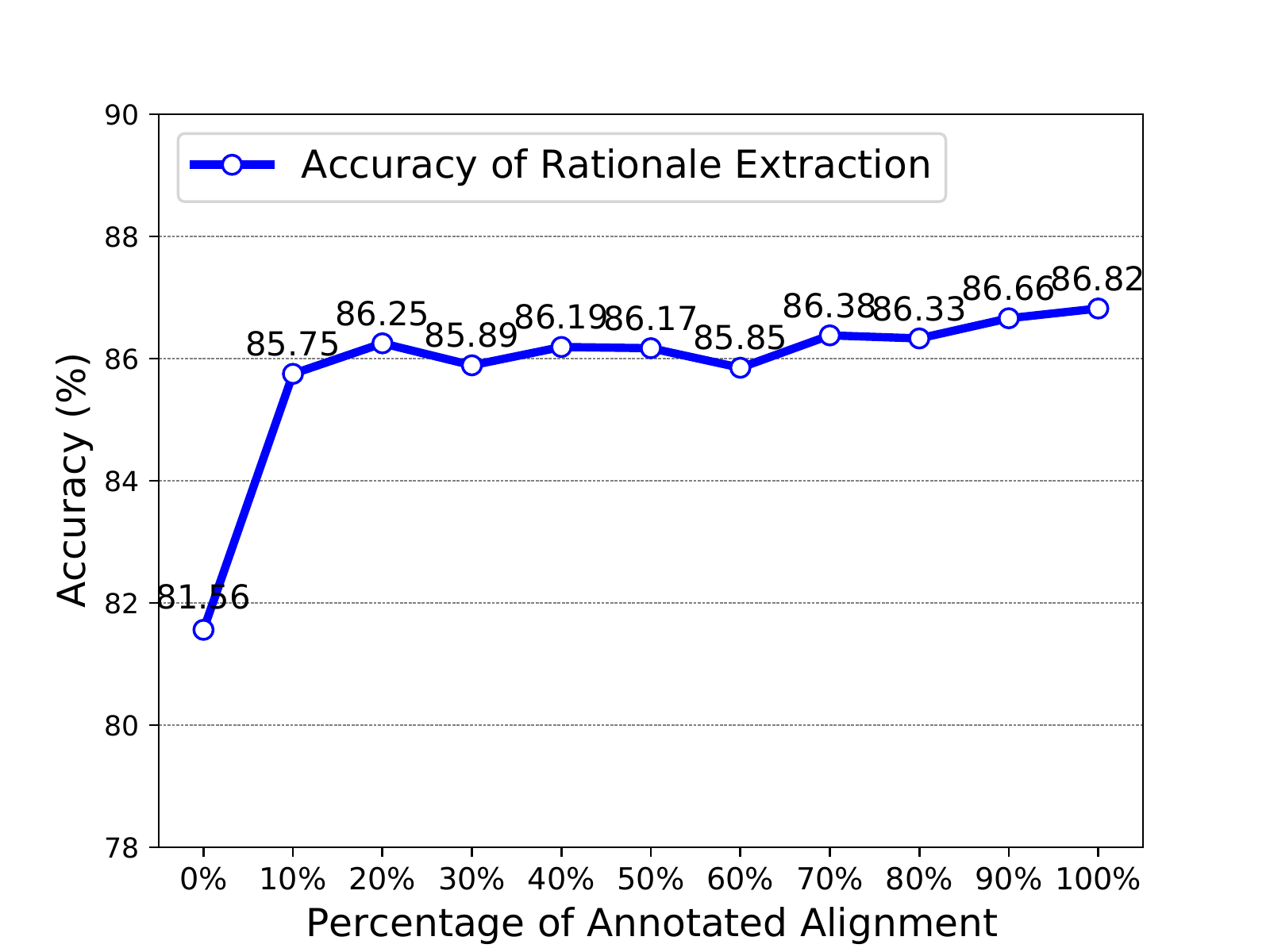}
    \caption{Rationale extraction accuracy of IOT-Match w.r.t. different percentages of labeled alignments.}
    \label{fig:semi}
\end{figure}
\subsection{Robustness under limited labels (RQ3)}
One advantage of IOT-Match is its capability of learning to extract the pro and con rationales from human-labeled rationale alignments in a semi-supervised manner, because,
in real practice, manually labeling the rationale alignments is expensive and time-consuming. 

We conduct experiments to test the rationale extraction accuracy w.r.t. different amounts of labeled alignments. 
Specifically, we configure IOT-Match to extract rationales given different ratios of labeled alignments $\hat{\mathbf{A}}$ in Eq.~\eqref{eq:IOT4Legal}
(from 0\% to 100\% where 0\% means no labels available,
and 100\% means fully supervised learning). 
Figure~\ref{fig:semi} illustrates the extraction accuracy w.r.t. the ratio of labeled alignments on ELAM data. 
We find that IOT-Match shows competitive performances when only 10\%\textasciitilde20\% of the labeled alignments are involved in learning. 
The results indicate that with only a small fraction of the alignment labels, IOT-Match can still learn the cross-case sentence-level affinity matrix $\mathbf{C}$ with high accuracy, and accurately extract the pro and con rationales. 

\begin{figure}
    \centering
    \subfigure[MLMC]{
    \includegraphics[width=0.45\linewidth]{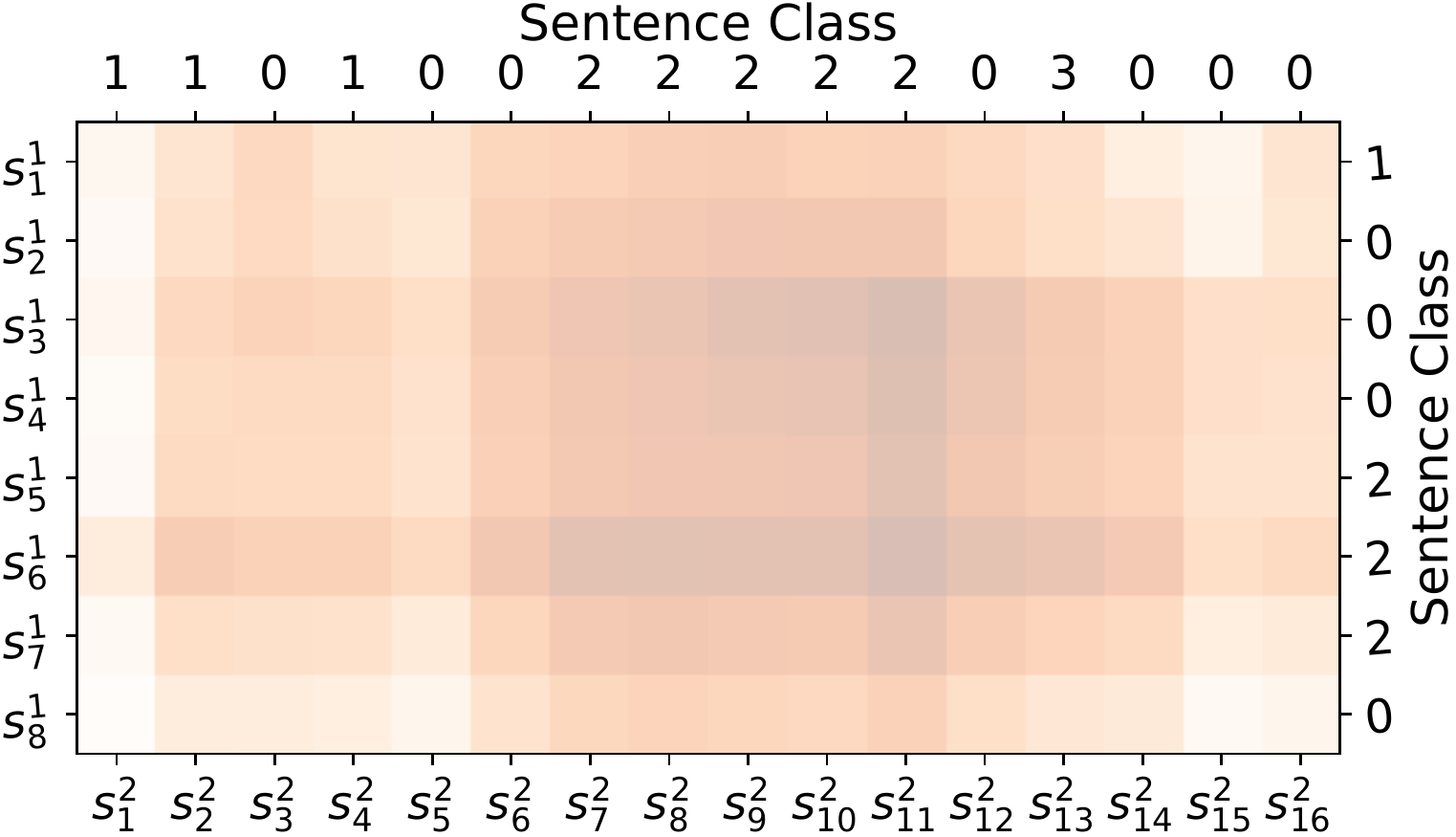}
    \label{fig:sub-a}
    }
    \subfigure[DecAtt]{
    \includegraphics[width=0.45\linewidth]{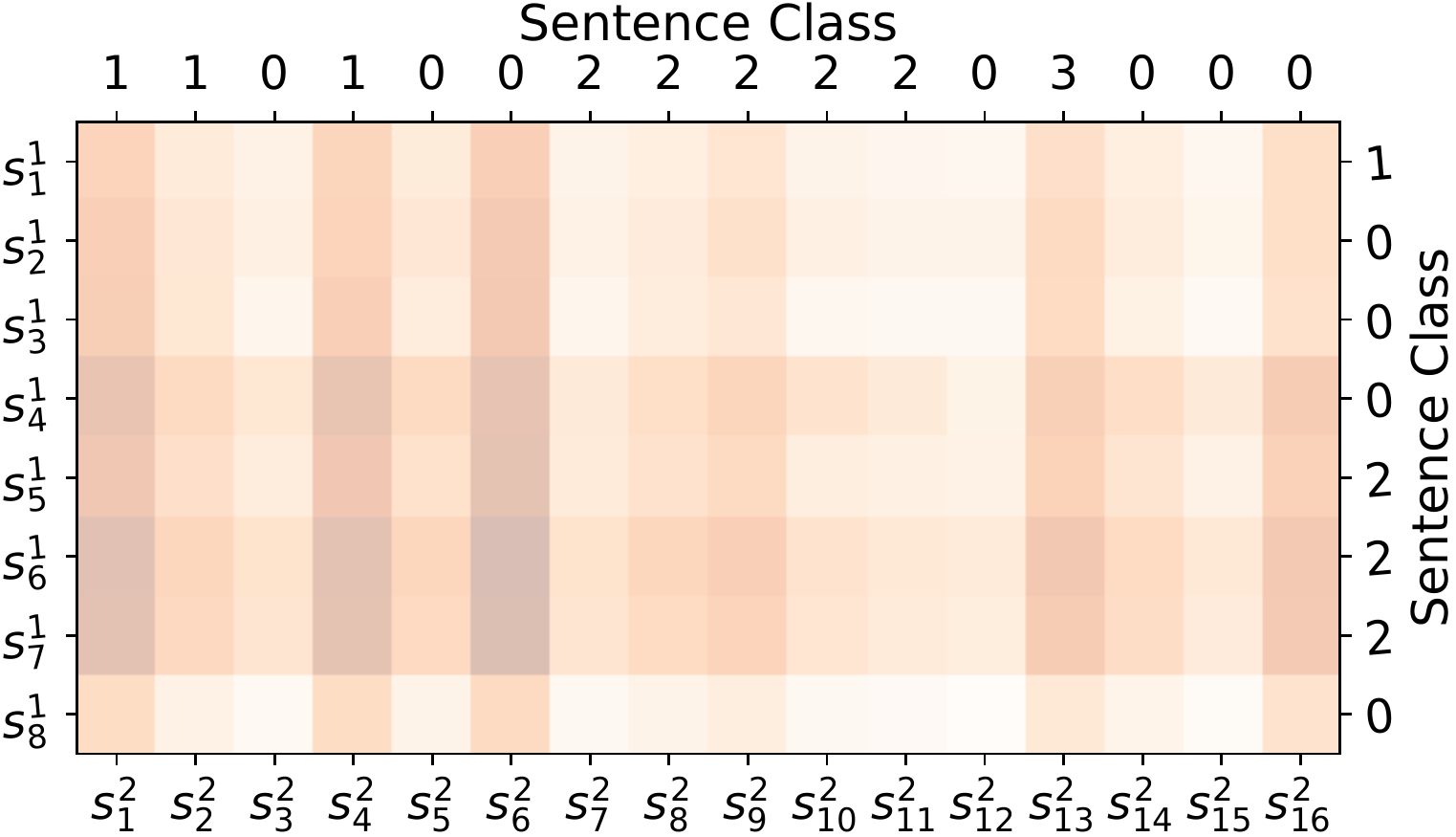}
    \label{fig:sub-b}
    }\\
    \subfigure[IOT-Match]{
    \includegraphics[width=0.45\linewidth]{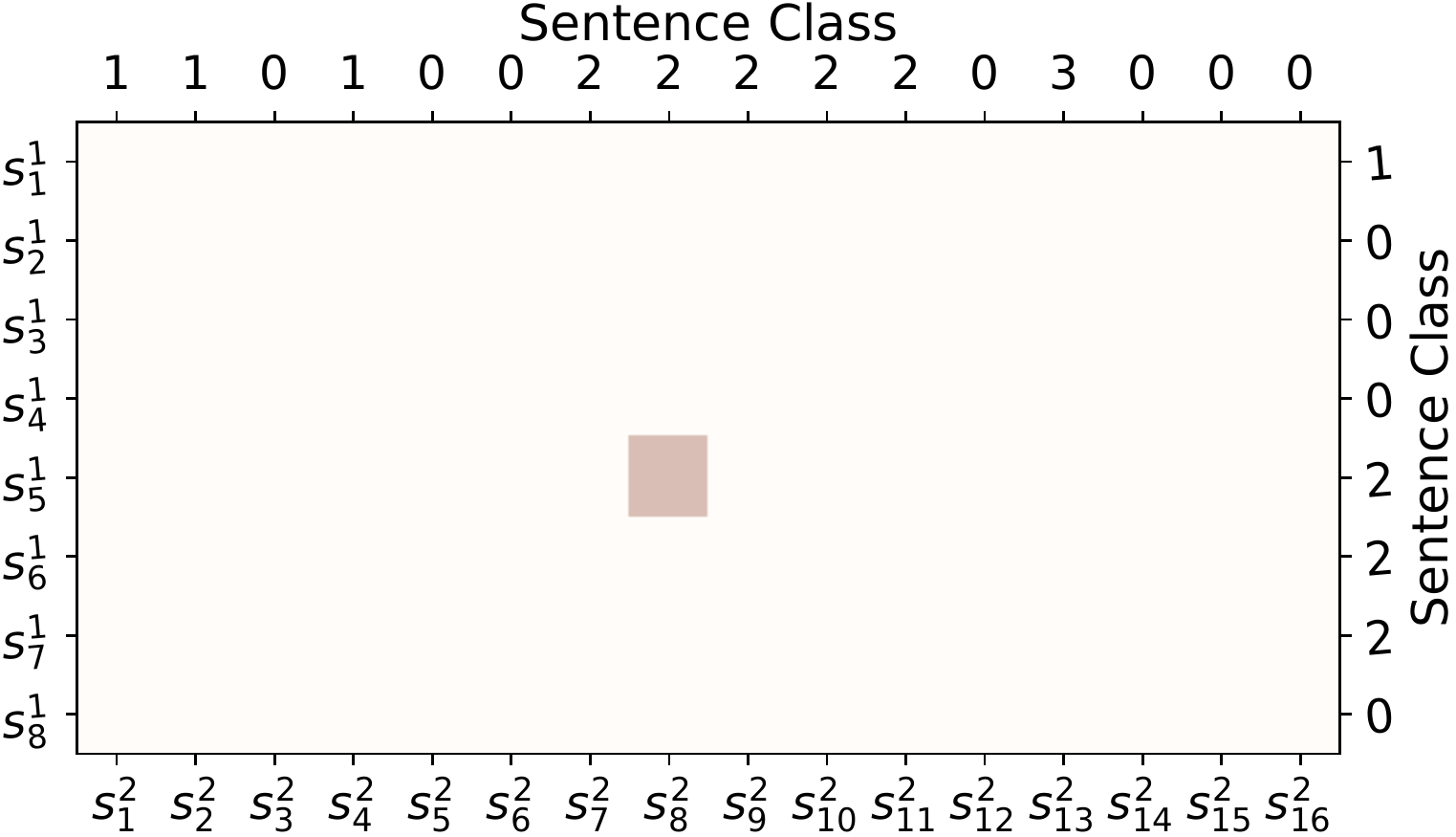}
    \label{fig:sub-c}
    }
    \subfigure[Human labeled alignments]{
    \includegraphics[width=0.45\linewidth]{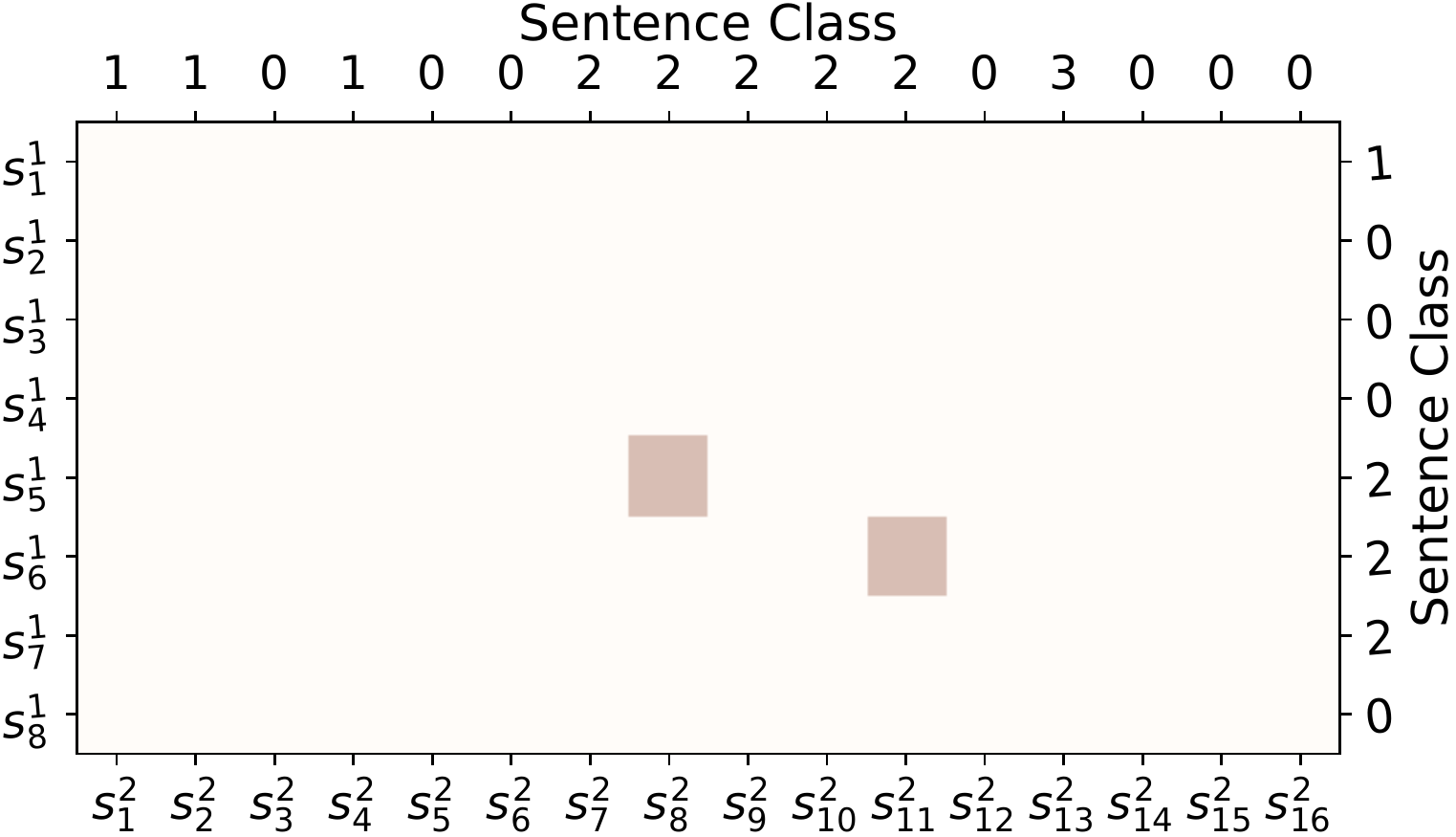}
    \label{fig:sub-d}
    }
    \subfigure[Aligned sentence pairs in the example cases (translated from Chinese)]{
    \includegraphics[width=\linewidth]{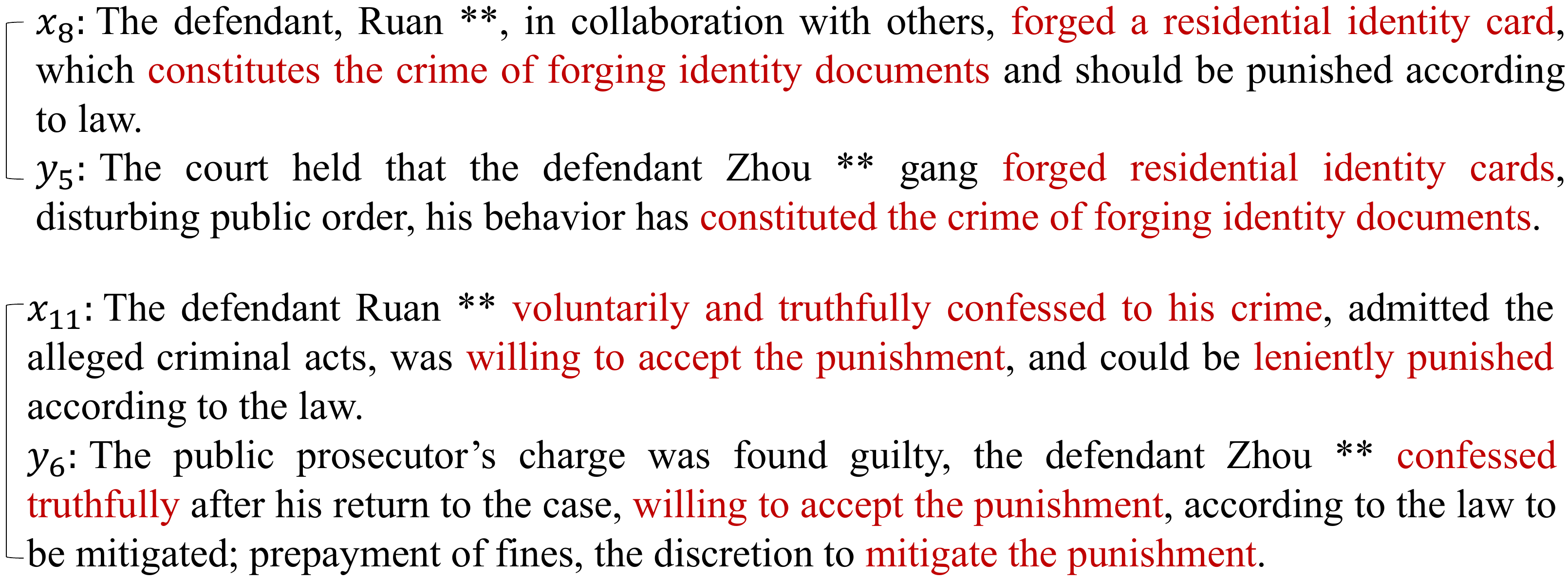}
    \label{fig:sub-e}
    }
    \caption{Rationale alignments of an example case pair from ELAM test set where 10\% of alignments are used for training. (a), (b), and (c): predicted rationale alignments; (d): human labeled alignments; (e): two labeled sentence pairs.
    }
    \label{fig:visual}
\end{figure}

Figure~\ref{fig:visual} shows the predicted rationale alignments for an example legal case pair (used in Figure~\ref{fig:exmaple}) from ELAM test set. 
The models are trained when only 10\% of the alignment labels are used. 
From the results, we find that MLMC and DecAtt output dense alignments (Figure~\ref{fig:visual}(a) and (b)), which are not accurate (ground-truth alignments are shown in Figure~\ref{fig:visual}(d)) and are hard to be explained. 
In contrast, IOT-Match not only accurately identifies and aligned the rationales (Figure~\ref{fig:visual}(c)), but also is explainable due to its sparseness.  
The results verify that IOT-Match is able to robustly and accurately extract and align the rationales when the alignment labels are insufficient. 
\section{Conclusion}
This paper proposes a novel inverse optimal transport-based model called IOT-Match for explainable legal case matching. 
IOT-Match is not only able to extract the pro and con rationales and generates natural language explanations for legal case matching, but is also robust to alignment label insufficiency. 
A new dataset is created to facilitate the study of explainable legal case matching.
Comprehensive experimental results showed that IOT-Match consistently outperformed state-of-the-art baselines in terms of matching accuracy. 
The empirical analysis verified that the extracted rationales and the generated explanations are not only consistent with human annotations but also faithful to the final matching prediction.
\begin{acks}
This work was funded by the National Key R\&D Program of China (2019YFE0198200), 
the National Natural Science Foundation of China (61872338, 61832017, 62106271, 62102420), Beijing Outstanding Young Scientist Program NO. BJJWZYJH012019100020098, the Mainland-Hong Kong Joint Funding Scheme (MHP/001/19) from the Innovation and Technology Commission (ITC) of Hong Kong, Intelligent Social Governance Interdisciplinary Platform, Major Innovation \& Planning Interdisciplinary Platform for the ``Double-First Class'' Initiative, Renmin University of China, and Public Policy and Decision-making Research Lab of Renmin University of China.
\end{acks}
\balance
\bibliographystyle{ACM-Reference-Format}
\bibliography{sample-base}


\begin{thebibliography}{55}


\ifx \showCODEN    \undefined \def \showCODEN     #1{\unskip}     \fi
\ifx \showDOI      \undefined \def \showDOI       #1{#1}\fi
\ifx \showISBNx    \undefined \def \showISBNx     #1{\unskip}     \fi
\ifx \showISBNxiii \undefined \def \showISBNxiii  #1{\unskip}     \fi
\ifx \showISSN     \undefined \def \showISSN      #1{\unskip}     \fi
\ifx \showLCCN     \undefined \def \showLCCN      #1{\unskip}     \fi
\ifx \shownote     \undefined \def \shownote      #1{#1}          \fi
\ifx \showarticletitle \undefined \def \showarticletitle #1{#1}   \fi
\ifx \showURL      \undefined \def \showURL       {\relax}        \fi
\providecommand\bibfield[2]{#2}
\providecommand\bibinfo[2]{#2}
\providecommand\natexlab[1]{#1}
\providecommand\showeprint[2][]{arXiv:#2}

\bibitem[Alvarez-Melis and Jaakkola(2018)]%
        {alvarez2018gromov}
\bibfield{author}{\bibinfo{person}{David Alvarez-Melis} {and}
  \bibinfo{person}{Tommi Jaakkola}.} \bibinfo{year}{2018}\natexlab{}.
\newblock \showarticletitle{Gromov-Wasserstein Alignment of Word Embedding
  Spaces}. In \bibinfo{booktitle}{\emph{Proceedings of the 2018 Conference on
  Empirical Methods in Natural Language Processing}}.
  \bibinfo{pages}{1881--1890}.
\newblock


\bibitem[Alvarez{-}Melis et~al\mbox{.}(2019)]%
        {alvarez2019towards}
\bibfield{author}{\bibinfo{person}{David Alvarez{-}Melis},
  \bibinfo{person}{Stefanie Jegelka}, {and} \bibinfo{person}{Tommi~S.
  Jaakkola}.} \bibinfo{year}{2019}\natexlab{}.
\newblock \showarticletitle{Towards Optimal Transport with Global Invariances}.
  In \bibinfo{booktitle}{\emph{The 22nd International Conference on Artificial
  Intelligence and Statistics, {AISTATS} 2019, 16-18 April 2019, Naha, Okinawa,
  Japan}} \emph{(\bibinfo{series}{Proceedings of Machine Learning Research},
  Vol.~\bibinfo{volume}{89})}. \bibinfo{publisher}{{PMLR}},
  \bibinfo{pages}{1870--1879}.
\newblock
\urldef\tempurl%
\url{http://proceedings.mlr.press/v89/alvarez-melis19a.html}
\showURL{%
\tempurl}


\bibitem[Atkinson et~al\mbox{.}(2020)]%
        {atkinson2020explanation}
\bibfield{author}{\bibinfo{person}{Katie Atkinson}, \bibinfo{person}{Trevor
  J.~M. Bench{-}Capon}, {and} \bibinfo{person}{Danushka Bollegala}.}
  \bibinfo{year}{2020}\natexlab{}.
\newblock \showarticletitle{Explanation in {AI} and law: Past, present and
  future}.
\newblock \bibinfo{journal}{\emph{Artif. Intell.}}  \bibinfo{volume}{289}
  (\bibinfo{year}{2020}), \bibinfo{pages}{103387}.
\newblock
\urldef\tempurl%
\url{https://doi.org/10.1016/j.artint.2020.103387}
\showURL{%
\tempurl}


\bibitem[Beltagy et~al\mbox{.}(2020)]%
        {beltagy2020longformer}
\bibfield{author}{\bibinfo{person}{Iz Beltagy}, \bibinfo{person}{Matthew~E
  Peters}, {and} \bibinfo{person}{Arman Cohan}.}
  \bibinfo{year}{2020}\natexlab{}.
\newblock \showarticletitle{Longformer: The long-document transformer}.
\newblock \bibinfo{journal}{\emph{ArXiv preprint}}
  \bibinfo{volume}{abs/2004.05150} (\bibinfo{year}{2020}).
\newblock
\urldef\tempurl%
\url{https://arxiv.org/abs/2004.05150}
\showURL{%
\tempurl}


\bibitem[Bench-Capon et~al\mbox{.}(2012)]%
        {bench2012history}
\bibfield{author}{\bibinfo{person}{Trevor Bench-Capon},
  \bibinfo{person}{Micha{\l} Araszkiewicz}, \bibinfo{person}{Kevin Ashley},
  \bibinfo{person}{Katie Atkinson}, \bibinfo{person}{Floris Bex},
  \bibinfo{person}{Filipe Borges}, \bibinfo{person}{Daniele Bourcier},
  \bibinfo{person}{Paul Bourgine}, \bibinfo{person}{Jack~G Conrad},
  \bibinfo{person}{Enrico Francesconi}, {et~al\mbox{.}}}
  \bibinfo{year}{2012}\natexlab{}.
\newblock \showarticletitle{A history of AI and Law in 50 papers: 25 years of
  the international conference on AI and Law}.
\newblock \bibinfo{journal}{\emph{Artificial Intelligence and Law}}
  \bibinfo{volume}{20}, \bibinfo{number}{3} (\bibinfo{year}{2012}),
  \bibinfo{pages}{215--319}.
\newblock


\bibitem[Bengio et~al\mbox{.}(2013)]%
        {bengio2013estimating}
\bibfield{author}{\bibinfo{person}{Yoshua Bengio}, \bibinfo{person}{Nicholas
  L{\'e}onard}, {and} \bibinfo{person}{Aaron Courville}.}
  \bibinfo{year}{2013}\natexlab{}.
\newblock \showarticletitle{Estimating or propagating gradients through
  stochastic neurons for conditional computation}.
\newblock \bibinfo{journal}{\emph{ArXiv preprint}}
  \bibinfo{volume}{abs/1308.3432} (\bibinfo{year}{2013}).
\newblock
\urldef\tempurl%
\url{https://arxiv.org/abs/1308.3432}
\showURL{%
\tempurl}


\bibitem[Bhattacharya et~al\mbox{.}(2020a)]%
        {bhattacharya2020hier}
\bibfield{author}{\bibinfo{person}{Paheli Bhattacharya},
  \bibinfo{person}{Kripabandhu Ghosh}, \bibinfo{person}{Arindam Pal}, {and}
  \bibinfo{person}{Saptarshi Ghosh}.} \bibinfo{year}{2020}\natexlab{a}.
\newblock \bibinfo{booktitle}{\emph{Hier-SPCNet: A Legal Statute
  Hierarchy-Based Heterogeneous Network for Computing Legal Case Document
  Similarity}}.
\newblock \bibinfo{publisher}{Association for Computing Machinery},
  \bibinfo{address}{New York, NY, USA}, \bibinfo{pages}{1657–1660}.
\newblock
\showISBNx{9781450380164}


\bibitem[Bhattacharya et~al\mbox{.}(2020b)]%
        {bhattacharya2020methods}
\bibfield{author}{\bibinfo{person}{Paheli Bhattacharya},
  \bibinfo{person}{Kripabandhu Ghosh}, \bibinfo{person}{Arindam Pal}, {and}
  \bibinfo{person}{Saptarshi Ghosh}.} \bibinfo{year}{2020}\natexlab{b}.
\newblock \showarticletitle{Methods for Computing Legal Document Similarity:
  {A} Comparative Study}.
\newblock \bibinfo{journal}{\emph{CoRR}}  \bibinfo{volume}{abs/2004.12307}
  (\bibinfo{year}{2020}).
\newblock
\showeprint[arXiv]{2004.12307}
\urldef\tempurl%
\url{https://arxiv.org/abs/2004.12307}
\showURL{%
\tempurl}


\bibitem[Bibal et~al\mbox{.}(2021)]%
        {bibal2021legal}
\bibfield{author}{\bibinfo{person}{Adrien Bibal}, \bibinfo{person}{Michael
  Lognoul}, \bibinfo{person}{Alexandre De~Streel}, {and}
  \bibinfo{person}{Beno{\^\i}t Fr{\'e}nay}.} \bibinfo{year}{2021}\natexlab{}.
\newblock \showarticletitle{Legal requirements on explainability in machine
  learning}.
\newblock \bibinfo{journal}{\emph{Artificial Intelligence and Law}}
  \bibinfo{volume}{29}, \bibinfo{number}{2} (\bibinfo{year}{2021}),
  \bibinfo{pages}{149--169}.
\newblock


\bibitem[Chakraborty et~al\mbox{.}(2020)]%
        {chakraborty2020hierarchical}
\bibfield{author}{\bibinfo{person}{Saptarshi Chakraborty},
  \bibinfo{person}{Debolina Paul}, {and} \bibinfo{person}{Swagatam Das}.}
  \bibinfo{year}{2020}\natexlab{}.
\newblock \showarticletitle{Hierarchical clustering with optimal transport}.
\newblock \bibinfo{journal}{\emph{Statistics \& Probability Letters}}
  \bibinfo{volume}{163} (\bibinfo{year}{2020}), \bibinfo{pages}{108781}.
\newblock


\bibitem[Chalkidis et~al\mbox{.}(2020)]%
        {chalkidis2020legal}
\bibfield{author}{\bibinfo{person}{Ilias Chalkidis}, \bibinfo{person}{Manos
  Fergadiotis}, \bibinfo{person}{Prodromos Malakasiotis},
  \bibinfo{person}{Nikolaos Aletras}, {and} \bibinfo{person}{Ion
  Androutsopoulos}.} \bibinfo{year}{2020}\natexlab{}.
\newblock \showarticletitle{{LEGAL-BERT:} "Preparing the Muppets for Court'"}.
  In \bibinfo{booktitle}{\emph{Findings of the Association for Computational
  Linguistics: {EMNLP} 2020, Online Event, 16-20 November 2020}}
  \emph{(\bibinfo{series}{Findings of {ACL}}, Vol.~\bibinfo{volume}{{EMNLP}
  2020})}. \bibinfo{publisher}{Association for Computational Linguistics},
  \bibinfo{pages}{2898--2904}.
\newblock
\urldef\tempurl%
\url{https://doi.org/10.18653/v1/2020.findings-emnlp.261}
\showURL{%
\tempurl}


\bibitem[Chen et~al\mbox{.}(2020a)]%
        {chen2020graph}
\bibfield{author}{\bibinfo{person}{Liqun Chen}, \bibinfo{person}{Zhe Gan},
  \bibinfo{person}{Yu Cheng}, \bibinfo{person}{Linjie Li},
  \bibinfo{person}{Lawrence Carin}, {and} \bibinfo{person}{Jingjing Liu}.}
  \bibinfo{year}{2020}\natexlab{a}.
\newblock \showarticletitle{Graph Optimal Transport for Cross-Domain
  Alignment}. In \bibinfo{booktitle}{\emph{Proceedings of the 37th
  International Conference on Machine Learning, {ICML} 2020, 13-18 July 2020,
  Virtual Event}} \emph{(\bibinfo{series}{Proceedings of Machine Learning
  Research}, Vol.~\bibinfo{volume}{119})}. \bibinfo{publisher}{{PMLR}},
  \bibinfo{pages}{1542--1553}.
\newblock
\urldef\tempurl%
\url{http://proceedings.mlr.press/v119/chen20e.html}
\showURL{%
\tempurl}


\bibitem[Chen et~al\mbox{.}(2019)]%
        {chen2018improving}
\bibfield{author}{\bibinfo{person}{Liqun Chen}, \bibinfo{person}{Yizhe Zhang},
  \bibinfo{person}{Ruiyi Zhang}, \bibinfo{person}{Chenyang Tao},
  \bibinfo{person}{Zhe Gan}, \bibinfo{person}{Haichao Zhang},
  \bibinfo{person}{Bai Li}, \bibinfo{person}{Dinghan Shen},
  \bibinfo{person}{Changyou Chen}, {and} \bibinfo{person}{Lawrence Carin}.}
  \bibinfo{year}{2019}\natexlab{}.
\newblock \showarticletitle{Improving Sequence-to-Sequence Learning via Optimal
  Transport}. In \bibinfo{booktitle}{\emph{7th International Conference on
  Learning Representations, {ICLR} 2019, New Orleans, LA, USA, May 6-9, 2019}}.
  \bibinfo{publisher}{OpenReview.net}.
\newblock
\urldef\tempurl%
\url{https://openreview.net/forum?id=S1xtAjR5tX}
\showURL{%
\tempurl}


\bibitem[Chen et~al\mbox{.}(2020b)]%
        {chen2020simple}
\bibfield{author}{\bibinfo{person}{Ting Chen}, \bibinfo{person}{Simon
  Kornblith}, \bibinfo{person}{Mohammad Norouzi}, {and}
  \bibinfo{person}{Geoffrey~E. Hinton}.} \bibinfo{year}{2020}\natexlab{b}.
\newblock \showarticletitle{A Simple Framework for Contrastive Learning of
  Visual Representations}. In \bibinfo{booktitle}{\emph{Proceedings of the 37th
  International Conference on Machine Learning, {ICML} 2020, 13-18 July 2020,
  Virtual Event}} \emph{(\bibinfo{series}{Proceedings of Machine Learning
  Research}, Vol.~\bibinfo{volume}{119})}. \bibinfo{publisher}{{PMLR}},
  \bibinfo{pages}{1597--1607}.
\newblock
\urldef\tempurl%
\url{http://proceedings.mlr.press/v119/chen20j.html}
\showURL{%
\tempurl}


\bibitem[Cheng et~al\mbox{.}(2020)]%
        {cheng2020argument}
\bibfield{author}{\bibinfo{person}{Liying Cheng}, \bibinfo{person}{Lidong
  Bing}, \bibinfo{person}{Qian Yu}, \bibinfo{person}{Wei Lu}, {and}
  \bibinfo{person}{Luo Si}.} \bibinfo{year}{2020}\natexlab{}.
\newblock \showarticletitle{{APE}: Argument Pair Extraction from Peer Review
  and Rebuttal via Multi-task Learning}. In
  \bibinfo{booktitle}{\emph{Proceedings of the 2020 Conference on Empirical
  Methods in Natural Language Processing (EMNLP)}}.
  \bibinfo{publisher}{Association for Computational Linguistics},
  \bibinfo{address}{Online}, \bibinfo{pages}{7000--7011}.
\newblock
\urldef\tempurl%
\url{https://aclanthology.org/2020.emnlp-main.569}
\showURL{%
\tempurl}


\bibitem[Cheng et~al\mbox{.}(2021)]%
        {cheng2021argument}
\bibfield{author}{\bibinfo{person}{Liying Cheng}, \bibinfo{person}{Tianyu Wu},
  \bibinfo{person}{Lidong Bing}, {and} \bibinfo{person}{Luo Si}.}
  \bibinfo{year}{2021}\natexlab{}.
\newblock \showarticletitle{Argument Pair Extraction via Attention-guided
  Multi-Layer Multi-Cross Encoding}. In \bibinfo{booktitle}{\emph{Proceedings
  of the 59th Annual Meeting of the Association for Computational Linguistics
  and the 11th International Joint Conference on Natural Language Processing
  (Volume 1: Long Papers)}}. \bibinfo{publisher}{Association for Computational
  Linguistics}, \bibinfo{address}{Online}, \bibinfo{pages}{6341--6353}.
\newblock
\urldef\tempurl%
\url{https://aclanthology.org/2021.acl-long.496}
\showURL{%
\tempurl}


\bibitem[Cuturi(2013)]%
        {cuturi2013sinkhorn}
\bibfield{author}{\bibinfo{person}{Marco Cuturi}.}
  \bibinfo{year}{2013}\natexlab{}.
\newblock \showarticletitle{Sinkhorn Distances: Lightspeed Computation of
  Optimal Transport}. In \bibinfo{booktitle}{\emph{Advances in Neural
  Information Processing Systems}}, Vol.~\bibinfo{volume}{26}.
  \bibinfo{publisher}{Curran Associates, Inc.}
\newblock
\urldef\tempurl%
\url{https://proceedings.neurips.cc/paper/2013/file/af21d0c97db2e27e13572cbf59eb343d-Paper.pdf}
\showURL{%
\tempurl}


\bibitem[DeYoung et~al\mbox{.}(2020)]%
        {deyoung2020eraser}
\bibfield{author}{\bibinfo{person}{Jay DeYoung}, \bibinfo{person}{Sarthak
  Jain}, \bibinfo{person}{Nazneen~Fatema Rajani}, \bibinfo{person}{Eric
  Lehman}, \bibinfo{person}{Caiming Xiong}, \bibinfo{person}{Richard Socher},
  {and} \bibinfo{person}{Byron~C. Wallace}.} \bibinfo{year}{2020}\natexlab{}.
\newblock \showarticletitle{{ERASER}: {A} Benchmark to Evaluate Rationalized
  {NLP} Models}. In \bibinfo{booktitle}{\emph{Proceedings of the 58th Annual
  Meeting of the Association for Computational Linguistics}}.
  \bibinfo{publisher}{Association for Computational Linguistics},
  \bibinfo{address}{Online}, \bibinfo{pages}{4443--4458}.
\newblock
\urldef\tempurl%
\url{https://aclanthology.org/2020.acl-main.408}
\showURL{%
\tempurl}


\bibitem[Doshi{-}Velez et~al\mbox{.}(2017)]%
        {doshi2017accountability}
\bibfield{author}{\bibinfo{person}{Finale Doshi{-}Velez},
  \bibinfo{person}{Mason Kortz}, \bibinfo{person}{Ryan Budish},
  \bibinfo{person}{Chris Bavitz}, \bibinfo{person}{Sam Gershman},
  \bibinfo{person}{David O'Brien}, \bibinfo{person}{Stuart Schieber},
  \bibinfo{person}{James Waldo}, \bibinfo{person}{David Weinberger}, {and}
  \bibinfo{person}{Alexandra Wood}.} \bibinfo{year}{2017}\natexlab{}.
\newblock \showarticletitle{Accountability of {AI} Under the Law: The Role of
  Explanation}.
\newblock \bibinfo{journal}{\emph{CoRR}}  \bibinfo{volume}{abs/1711.01134}
  (\bibinfo{year}{2017}).
\newblock
\showeprint[arXiv]{1711.01134}
\urldef\tempurl%
\url{http://arxiv.org/abs/1711.01134}
\showURL{%
\tempurl}


\bibitem[Dupuy et~al\mbox{.}(2016)]%
        {dupuy2016estimating}
\bibfield{author}{\bibinfo{person}{Arnaud Dupuy}, \bibinfo{person}{Alfred
  Galichon}, {and} \bibinfo{person}{Yifei Sun}.}
  \bibinfo{year}{2016}\natexlab{}.
\newblock \showarticletitle{Estimating matching affinity matrix under low-rank
  constraints}.
\newblock \bibinfo{journal}{\emph{ArXiv preprint}}
  \bibinfo{volume}{abs/1612.09585} (\bibinfo{year}{2016}).
\newblock
\urldef\tempurl%
\url{https://arxiv.org/abs/1612.09585}
\showURL{%
\tempurl}


\bibitem[Gao et~al\mbox{.}(2021)]%
        {gao2021simcse}
\bibfield{author}{\bibinfo{person}{Tianyu Gao}, \bibinfo{person}{Xingcheng
  Yao}, {and} \bibinfo{person}{Danqi Chen}.} \bibinfo{year}{2021}\natexlab{}.
\newblock \showarticletitle{SimCSE: Simple Contrastive Learning of Sentence
  Embeddings}. In \bibinfo{booktitle}{\emph{Proceedings of the 2021 Conference
  on Empirical Methods in Natural Language Processing, {EMNLP} 2021, Virtual
  Event / Punta Cana, Dominican Republic, 7-11 November, 2021}}.
  \bibinfo{publisher}{Association for Computational Linguistics},
  \bibinfo{pages}{6894--6910}.
\newblock
\urldef\tempurl%
\url{https://doi.org/10.18653/v1/2021.emnlp-main.552}
\showURL{%
\tempurl}


\bibitem[Gehring et~al\mbox{.}(2017)]%
        {gehring2017convolutional}
\bibfield{author}{\bibinfo{person}{Jonas Gehring}, \bibinfo{person}{Michael
  Auli}, \bibinfo{person}{David Grangier}, \bibinfo{person}{Denis Yarats},
  {and} \bibinfo{person}{Yann~N. Dauphin}.} \bibinfo{year}{2017}\natexlab{}.
\newblock \showarticletitle{Convolutional Sequence to Sequence Learning}. In
  \bibinfo{booktitle}{\emph{Proceedings of the 34th International Conference on
  Machine Learning, {ICML} 2017, Sydney, NSW, Australia, 6-11 August 2017}}
  \emph{(\bibinfo{series}{Proceedings of Machine Learning Research},
  Vol.~\bibinfo{volume}{70})}. \bibinfo{publisher}{{PMLR}},
  \bibinfo{pages}{1243--1252}.
\newblock
\urldef\tempurl%
\url{http://proceedings.mlr.press/v70/gehring17a.html}
\showURL{%
\tempurl}


\bibitem[Jiang et~al\mbox{.}(2018)]%
        {jiang2018interpretable}
\bibfield{author}{\bibinfo{person}{Xin Jiang}, \bibinfo{person}{Hai Ye},
  \bibinfo{person}{Zhunchen Luo}, \bibinfo{person}{WenHan Chao}, {and}
  \bibinfo{person}{Wenjia Ma}.} \bibinfo{year}{2018}\natexlab{}.
\newblock \showarticletitle{Interpretable Rationale Augmented Charge Prediction
  System}. In \bibinfo{booktitle}{\emph{Proceedings of the 27th International
  Conference on Computational Linguistics: System Demonstrations}}.
  \bibinfo{publisher}{Association for Computational Linguistics},
  \bibinfo{address}{Santa Fe, New Mexico}, \bibinfo{pages}{146--151}.
\newblock
\urldef\tempurl%
\url{https://aclanthology.org/C18-2032}
\showURL{%
\tempurl}


\bibitem[Kingma and Ba(2015)]%
        {kingma2014adam}
\bibfield{author}{\bibinfo{person}{Diederik~P. Kingma} {and}
  \bibinfo{person}{Jimmy Ba}.} \bibinfo{year}{2015}\natexlab{}.
\newblock \showarticletitle{Adam: {A} Method for Stochastic Optimization}. In
  \bibinfo{booktitle}{\emph{3rd International Conference on Learning
  Representations, {ICLR} 2015, San Diego, CA, USA, May 7-9, 2015, Conference
  Track Proceedings}}.
\newblock
\urldef\tempurl%
\url{http://arxiv.org/abs/1412.6980}
\showURL{%
\tempurl}


\bibitem[Kumar et~al\mbox{.}(2011)]%
        {kumar2011similarity}
\bibfield{author}{\bibinfo{person}{Sushanta Kumar}, \bibinfo{person}{P.~Krishna
  Reddy}, \bibinfo{person}{V.~Balakista Reddy}, {and} \bibinfo{person}{Aditya
  Singh}.} \bibinfo{year}{2011}\natexlab{}.
\newblock \showarticletitle{Similarity analysis of legal judgments}. In
  \bibinfo{booktitle}{\emph{Proceedings of the 4th Bangalore Annual Compute
  Conference, Compute 2011, Bangalore, India, March 25-26, 2011}}.
  \bibinfo{publisher}{{ACM}}, \bibinfo{pages}{17}.
\newblock
\urldef\tempurl%
\url{https://doi.org/10.1145/1980422.1980439}
\showURL{%
\tempurl}


\bibitem[Kumar and Talukdar(2020)]%
        {kumar2020nile}
\bibfield{author}{\bibinfo{person}{Sawan Kumar} {and} \bibinfo{person}{Partha
  Talukdar}.} \bibinfo{year}{2020}\natexlab{}.
\newblock \showarticletitle{{NILE} : Natural Language Inference with Faithful
  Natural Language Explanations}. In \bibinfo{booktitle}{\emph{Proceedings of
  the 58th Annual Meeting of the Association for Computational Linguistics}}.
  \bibinfo{publisher}{Association for Computational Linguistics},
  \bibinfo{address}{Online}, \bibinfo{pages}{8730--8742}.
\newblock
\urldef\tempurl%
\url{https://aclanthology.org/2020.acl-main.771}
\showURL{%
\tempurl}


\bibitem[Li et~al\mbox{.}(2019)]%
        {li2019learning}
\bibfield{author}{\bibinfo{person}{Ruilin Li}, \bibinfo{person}{Xiaojing Ye},
  \bibinfo{person}{Haomin Zhou}, {and} \bibinfo{person}{Hongyuan Zha}.}
  \bibinfo{year}{2019}\natexlab{}.
\newblock \showarticletitle{Learning to Match via Inverse Optimal Transport}.
\newblock \bibinfo{journal}{\emph{J. Mach. Learn. Res.}}  \bibinfo{volume}{20}
  (\bibinfo{year}{2019}), \bibinfo{pages}{80:1--80:37}.
\newblock
\urldef\tempurl%
\url{http://jmlr.org/papers/v20/18-700.html}
\showURL{%
\tempurl}


\bibitem[Liu et~al\mbox{.}(2021)]%
        {liu2021interpretable}
\bibfield{author}{\bibinfo{person}{Liting Liu}, \bibinfo{person}{Wenzheng
  Zhang}, \bibinfo{person}{Jie Liu}, \bibinfo{person}{Wenxuan Shi}, {and}
  \bibinfo{person}{Yalou Huang}.} \bibinfo{year}{2021}\natexlab{}.
\newblock \showarticletitle{Interpretable Charge Prediction for Legal Cases
  based on Interdependent Legal Information}. In
  \bibinfo{booktitle}{\emph{International Joint Conference on Neural Networks,
  {IJCNN} 2021, Shenzhen, China, July 18-22, 2021}}.
  \bibinfo{publisher}{{IEEE}}, \bibinfo{pages}{1--8}.
\newblock
\urldef\tempurl%
\url{https://doi.org/10.1109/IJCNN52387.2021.9533902}
\showURL{%
\tempurl}


\bibitem[Liu and Liu(2021)]%
        {liu2021simcls}
\bibfield{author}{\bibinfo{person}{Yixin Liu} {and} \bibinfo{person}{Pengfei
  Liu}.} \bibinfo{year}{2021}\natexlab{}.
\newblock \showarticletitle{{S}im{CLS}: A Simple Framework for Contrastive
  Learning of Abstractive Summarization}. In
  \bibinfo{booktitle}{\emph{Proceedings of the 59th Annual Meeting of the
  Association for Computational Linguistics and the 11th International Joint
  Conference on Natural Language Processing (Volume 2: Short Papers)}}.
  \bibinfo{publisher}{Association for Computational Linguistics},
  \bibinfo{address}{Online}, \bibinfo{pages}{1065--1072}.
\newblock
\urldef\tempurl%
\url{https://aclanthology.org/2021.acl-short.135}
\showURL{%
\tempurl}


\bibitem[Luu et~al\mbox{.}(2021)]%
        {luu2021explaining}
\bibfield{author}{\bibinfo{person}{Kelvin Luu}, \bibinfo{person}{Xinyi Wu},
  \bibinfo{person}{Rik Koncel-Kedziorski}, \bibinfo{person}{Kyle Lo},
  \bibinfo{person}{Isabel Cachola}, {and} \bibinfo{person}{Noah~A. Smith}.}
  \bibinfo{year}{2021}\natexlab{}.
\newblock \showarticletitle{Explaining Relationships Between Scientific
  Documents}. In \bibinfo{booktitle}{\emph{Proceedings of the 59th Annual
  Meeting of the Association for Computational Linguistics and the 11th
  International Joint Conference on Natural Language Processing (Volume 1: Long
  Papers)}}. \bibinfo{publisher}{Association for Computational Linguistics},
  \bibinfo{address}{Online}, \bibinfo{pages}{2130--2144}.
\newblock
\urldef\tempurl%
\url{https://aclanthology.org/2021.acl-long.166}
\showURL{%
\tempurl}


\bibitem[Ma et~al\mbox{.}(2021)]%
        {ma2021lecard}
\bibfield{author}{\bibinfo{person}{Yixiao Ma}, \bibinfo{person}{Yunqiu Shao},
  \bibinfo{person}{Yueyue Wu}, \bibinfo{person}{Yiqun Liu},
  \bibinfo{person}{Ruizhe Zhang}, \bibinfo{person}{Min Zhang}, {and}
  \bibinfo{person}{Shaoping Ma}.} \bibinfo{year}{2021}\natexlab{}.
\newblock \showarticletitle{LeCaRD: A Legal Case Retrieval Dataset for Chinese
  Law System}.
\newblock \bibinfo{journal}{\emph{Information Retrieval (IR)}}
  \bibinfo{volume}{2} (\bibinfo{year}{2021}), \bibinfo{pages}{22}.
\newblock


\bibitem[Minocha et~al\mbox{.}(2015)]%
        {minocha2015finding}
\bibfield{author}{\bibinfo{person}{Akshay Minocha}, \bibinfo{person}{Navjyoti
  Singh}, {and} \bibinfo{person}{Arjit Srivastava}.}
  \bibinfo{year}{2015}\natexlab{}.
\newblock \showarticletitle{Finding Relevant Indian Judgments Using Dispersion
  of Citation Network}. In \bibinfo{booktitle}{\emph{Proceedings of the 24th
  International Conference on World Wide Web}} (Florence, Italy)
  \emph{(\bibinfo{series}{WWW '15 Companion})}. \bibinfo{publisher}{Association
  for Computing Machinery}, \bibinfo{address}{New York, NY, USA},
  \bibinfo{pages}{1085–1088}.
\newblock
\showISBNx{9781450334730}
\urldef\tempurl%
\url{https://doi.org/10.1145/2740908.2744717}
\showURL{%
\tempurl}


\bibitem[Monroy et~al\mbox{.}(2013)]%
        {monroy2013link}
\bibfield{author}{\bibinfo{person}{Alfredo~L{\'{o}}pez Monroy},
  \bibinfo{person}{Hiram Calvo}, \bibinfo{person}{Alexander~F. Gelbukh}, {and}
  \bibinfo{person}{Georgina~Garc{\'{\i}}a Pacheco}.}
  \bibinfo{year}{2013}\natexlab{}.
\newblock \showarticletitle{Link Analysis for Representing and Retrieving Legal
  Information}. In \bibinfo{booktitle}{\emph{Computational Linguistics and
  Intelligent Text Processing - 14th International Conference, CICLing 2013,
  Samos, Greece, March 24-30, 2013, Proceedings, Part {II}}}
  \emph{(\bibinfo{series}{Lecture Notes in Computer Science},
  Vol.~\bibinfo{volume}{7817})}. \bibinfo{publisher}{Springer},
  \bibinfo{pages}{380--393}.
\newblock
\urldef\tempurl%
\url{https://doi.org/10.1007/978-3-642-37256-8\_32}
\showURL{%
\tempurl}


\bibitem[Paranjape et~al\mbox{.}(2020)]%
        {paranjape2020information}
\bibfield{author}{\bibinfo{person}{Bhargavi Paranjape}, \bibinfo{person}{Mandar
  Joshi}, \bibinfo{person}{John Thickstun}, \bibinfo{person}{Hannaneh
  Hajishirzi}, {and} \bibinfo{person}{Luke Zettlemoyer}.}
  \bibinfo{year}{2020}\natexlab{}.
\newblock \showarticletitle{An Information Bottleneck Approach for Controlling
  Conciseness in Rationale Extraction}. In
  \bibinfo{booktitle}{\emph{Proceedings of the 2020 Conference on Empirical
  Methods in Natural Language Processing (EMNLP)}}.
  \bibinfo{publisher}{Association for Computational Linguistics},
  \bibinfo{address}{Online}, \bibinfo{pages}{1938--1952}.
\newblock
\urldef\tempurl%
\url{https://aclanthology.org/2020.emnlp-main.153}
\showURL{%
\tempurl}


\bibitem[Parikh et~al\mbox{.}(2016)]%
        {parikh2016decomposable}
\bibfield{author}{\bibinfo{person}{Ankur Parikh}, \bibinfo{person}{Oscar
  T{\"a}ckstr{\"o}m}, \bibinfo{person}{Dipanjan Das}, {and}
  \bibinfo{person}{Jakob Uszkoreit}.} \bibinfo{year}{2016}\natexlab{}.
\newblock \showarticletitle{A Decomposable Attention Model for Natural Language
  Inference}. In \bibinfo{booktitle}{\emph{Proceedings of the 2016 Conference
  on Empirical Methods in Natural Language Processing}}.
  \bibinfo{publisher}{Association for Computational Linguistics},
  \bibinfo{address}{Austin, Texas}, \bibinfo{pages}{2249--2255}.
\newblock
\urldef\tempurl%
\url{https://aclanthology.org/D16-1244}
\showURL{%
\tempurl}


\bibitem[Peyr{\'e} et~al\mbox{.}(2019)]%
        {peyre2019computational}
\bibfield{author}{\bibinfo{person}{Gabriel Peyr{\'e}}, \bibinfo{person}{Marco
  Cuturi}, {et~al\mbox{.}}} \bibinfo{year}{2019}\natexlab{}.
\newblock \showarticletitle{Computational optimal transport: With applications
  to data science}.
\newblock \bibinfo{journal}{\emph{Foundations and Trends{\textregistered} in
  Machine Learning}} \bibinfo{volume}{11}, \bibinfo{number}{5-6}
  (\bibinfo{year}{2019}), \bibinfo{pages}{355--607}.
\newblock


\bibitem[Raffel et~al\mbox{.}(2020)]%
        {raffel2019exploring}
\bibfield{author}{\bibinfo{person}{Colin Raffel}, \bibinfo{person}{Noam
  Shazeer}, \bibinfo{person}{Adam Roberts}, \bibinfo{person}{Katherine Lee},
  \bibinfo{person}{Sharan Narang}, \bibinfo{person}{Michael Matena},
  \bibinfo{person}{Yanqi Zhou}, \bibinfo{person}{Wei Li}, {and}
  \bibinfo{person}{Peter~J. Liu}.} \bibinfo{year}{2020}\natexlab{}.
\newblock \showarticletitle{Exploring the Limits of Transfer Learning with a
  Unified Text-to-Text Transformer}.
\newblock \bibinfo{journal}{\emph{J. Mach. Learn. Res.}}  \bibinfo{volume}{21}
  (\bibinfo{year}{2020}), \bibinfo{pages}{140:1--140:67}.
\newblock
\urldef\tempurl%
\url{http://jmlr.org/papers/v21/20-074.html}
\showURL{%
\tempurl}


\bibitem[Reimers and Gurevych(2019)]%
        {reimers2019sentence}
\bibfield{author}{\bibinfo{person}{Nils Reimers} {and} \bibinfo{person}{Iryna
  Gurevych}.} \bibinfo{year}{2019}\natexlab{}.
\newblock \showarticletitle{Sentence-{BERT}: Sentence Embeddings using
  {S}iamese {BERT}-Networks}. In \bibinfo{booktitle}{\emph{Proceedings of the
  2019 Conference on Empirical Methods in Natural Language Processing and the
  9th International Joint Conference on Natural Language Processing
  (EMNLP-IJCNLP)}}. \bibinfo{publisher}{Association for Computational
  Linguistics}, \bibinfo{address}{Hong Kong, China},
  \bibinfo{pages}{3982--3992}.
\newblock
\urldef\tempurl%
\url{https://aclanthology.org/D19-1410}
\showURL{%
\tempurl}


\bibitem[Saravanan et~al\mbox{.}(2009)]%
        {saravanan2009improving}
\bibfield{author}{\bibinfo{person}{Manavalan Saravanan},
  \bibinfo{person}{Balaraman Ravindran}, {and} \bibinfo{person}{Shivani
  Raman}.} \bibinfo{year}{2009}\natexlab{}.
\newblock \showarticletitle{Improving legal information retrieval using an
  ontological framework}.
\newblock \bibinfo{journal}{\emph{Artificial Intelligence and Law}}
  \bibinfo{volume}{17}, \bibinfo{number}{2} (\bibinfo{year}{2009}),
  \bibinfo{pages}{101--124}.
\newblock


\bibitem[Sha et~al\mbox{.}(2021)]%
        {sha2021learning}
\bibfield{author}{\bibinfo{person}{Lei Sha}, \bibinfo{person}{Oana{-}Maria
  Camburu}, {and} \bibinfo{person}{Thomas Lukasiewicz}.}
  \bibinfo{year}{2021}\natexlab{}.
\newblock \showarticletitle{Learning from the Best: Rationalizing Predictions
  by Adversarial Information Calibration}. In
  \bibinfo{booktitle}{\emph{Thirty-Fifth {AAAI} Conference on Artificial
  Intelligence, {AAAI} 2021, Virtual Event, February 2-9, 2021}}.
  \bibinfo{publisher}{{AAAI} Press}, \bibinfo{pages}{13771--13779}.
\newblock
\urldef\tempurl%
\url{https://ojs.aaai.org/index.php/AAAI/article/view/17623}
\showURL{%
\tempurl}


\bibitem[Shao et~al\mbox{.}(2020)]%
        {shao2020bert}
\bibfield{author}{\bibinfo{person}{Yunqiu Shao}, \bibinfo{person}{Jiaxin Mao},
  \bibinfo{person}{Yiqun Liu}, \bibinfo{person}{Weizhi Ma},
  \bibinfo{person}{Ken Satoh}, \bibinfo{person}{Min Zhang}, {and}
  \bibinfo{person}{Shaoping Ma}.} \bibinfo{year}{2020}\natexlab{}.
\newblock \showarticletitle{{BERT-PLI:} Modeling Paragraph-Level Interactions
  for Legal Case Retrieval}. In \bibinfo{booktitle}{\emph{Proceedings of the
  Twenty-Ninth International Joint Conference on Artificial Intelligence,
  {IJCAI} 2020}}. \bibinfo{publisher}{ijcai.org}, \bibinfo{pages}{3501--3507}.
\newblock
\urldef\tempurl%
\url{https://doi.org/10.24963/ijcai.2020/484}
\showURL{%
\tempurl}


\bibitem[Su(2021)]%
        {zhuiyit5pegasus}
\bibfield{author}{\bibinfo{person}{Jianlin Su}.}
  \bibinfo{year}{2021}\natexlab{}.
\newblock \bibinfo{booktitle}{\emph{T5 PEGASUS - ZhuiyiAI}}.
\newblock \bibinfo{type}{{T}echnical {R}eport}.
\newblock


\bibitem[Villani(2009)]%
        {villani2009optimal}
\bibfield{author}{\bibinfo{person}{C{\'e}dric Villani}.}
  \bibinfo{year}{2009}\natexlab{}.
\newblock \bibinfo{booktitle}{\emph{Optimal transport: old and new}}.
  Vol.~\bibinfo{volume}{338}.
\newblock \bibinfo{publisher}{Springer}.
\newblock


\bibitem[Xiao et~al\mbox{.}(2021)]%
        {xiao2021lawformer}
\bibfield{author}{\bibinfo{person}{Chaojun Xiao}, \bibinfo{person}{Xueyu Hu},
  \bibinfo{person}{Zhiyuan Liu}, \bibinfo{person}{Cunchao Tu}, {and}
  \bibinfo{person}{Maosong Sun}.} \bibinfo{year}{2021}\natexlab{}.
\newblock \showarticletitle{Lawformer: {A} Pre-trained Language Model for
  Chinese Legal Long Documents}.
\newblock \bibinfo{journal}{\emph{CoRR}}  \bibinfo{volume}{abs/2105.03887}
  (\bibinfo{year}{2021}).
\newblock
\showeprint[arXiv]{2105.03887}
\urldef\tempurl%
\url{https://arxiv.org/abs/2105.03887}
\showURL{%
\tempurl}


\bibitem[Xu et~al\mbox{.}(2019a)]%
        {xu2019scalable}
\bibfield{author}{\bibinfo{person}{Hongteng Xu}, \bibinfo{person}{Dixin Luo},
  {and} \bibinfo{person}{Lawrence Carin}.} \bibinfo{year}{2019}\natexlab{a}.
\newblock \showarticletitle{Scalable Gromov-Wasserstein Learning for Graph
  Partitioning and Matching}. In \bibinfo{booktitle}{\emph{Advances in Neural
  Information Processing Systems}}, Vol.~\bibinfo{volume}{32}.
  \bibinfo{publisher}{Curran Associates, Inc.}
\newblock
\urldef\tempurl%
\url{https://proceedings.neurips.cc/paper/2019/file/6e62a992c676f611616097dbea8ea030-Paper.pdf}
\showURL{%
\tempurl}


\bibitem[Xu et~al\mbox{.}(2019b)]%
        {xu2019gromov}
\bibfield{author}{\bibinfo{person}{Hongteng Xu}, \bibinfo{person}{Dixin Luo},
  \bibinfo{person}{Hongyuan Zha}, {and} \bibinfo{person}{Lawrence Carin}.}
  \bibinfo{year}{2019}\natexlab{b}.
\newblock \showarticletitle{Gromov-Wasserstein Learning for Graph Matching and
  Node Embedding}. In \bibinfo{booktitle}{\emph{Proceedings of the 36th
  International Conference on Machine Learning, {ICML} 2019, 9-15 June 2019,
  Long Beach, California, {USA}}} \emph{(\bibinfo{series}{Proceedings of
  Machine Learning Research}, Vol.~\bibinfo{volume}{97})}.
  \bibinfo{publisher}{{PMLR}}, \bibinfo{pages}{6932--6941}.
\newblock
\urldef\tempurl%
\url{http://proceedings.mlr.press/v97/xu19b.html}
\showURL{%
\tempurl}


\bibitem[Xu et~al\mbox{.}(2018)]%
        {xu2018distilled}
\bibfield{author}{\bibinfo{person}{Hongteng Xu}, \bibinfo{person}{Wenlin Wang},
  \bibinfo{person}{Wei Liu}, {and} \bibinfo{person}{Lawrence Carin}.}
  \bibinfo{year}{2018}\natexlab{}.
\newblock \showarticletitle{Distilled {W}asserstein learning for word embedding
  and topic modeling}.
\newblock \bibinfo{journal}{\emph{Advances in Neural Information Processing
  Systems}}  \bibinfo{volume}{31} (\bibinfo{year}{2018}).
\newblock


\bibitem[Ye et~al\mbox{.}(2018)]%
        {ye2018interpretable}
\bibfield{author}{\bibinfo{person}{Hai Ye}, \bibinfo{person}{Xin Jiang},
  \bibinfo{person}{Zhunchen Luo}, {and} \bibinfo{person}{Wenhan Chao}.}
  \bibinfo{year}{2018}\natexlab{}.
\newblock \showarticletitle{Interpretable Charge Predictions for Criminal
  Cases: Learning to Generate Court Views from Fact Descriptions}. In
  \bibinfo{booktitle}{\emph{Proceedings of the 2018 Conference of the North
  {A}merican Chapter of the Association for Computational Linguistics: Human
  Language Technologies, Volume 1 (Long Papers)}}.
  \bibinfo{publisher}{Association for Computational Linguistics},
  \bibinfo{address}{New Orleans, Louisiana}, \bibinfo{pages}{1854--1864}.
\newblock
\urldef\tempurl%
\url{https://aclanthology.org/N18-1168}
\showURL{%
\tempurl}


\bibitem[Yu and Koltun(2016)]%
        {yu2015multi}
\bibfield{author}{\bibinfo{person}{Fisher Yu} {and} \bibinfo{person}{Vladlen
  Koltun}.} \bibinfo{year}{2016}\natexlab{}.
\newblock \showarticletitle{Multi-Scale Context Aggregation by Dilated
  Convolutions}. In \bibinfo{booktitle}{\emph{4th International Conference on
  Learning Representations, {ICLR} 2016, San Juan, Puerto Rico, May 2-4, 2016,
  Conference Track Proceedings}}.
\newblock
\urldef\tempurl%
\url{http://arxiv.org/abs/1511.07122}
\showURL{%
\tempurl}


\bibitem[Yu et~al\mbox{.}(2020)]%
        {yu-etal-2020-wasserstein}
\bibfield{author}{\bibinfo{person}{Weijie Yu}, \bibinfo{person}{Chen Xu},
  \bibinfo{person}{Jun Xu}, \bibinfo{person}{Liang Pang},
  \bibinfo{person}{Xiaopeng Gao}, \bibinfo{person}{Xiaozhao Wang}, {and}
  \bibinfo{person}{Ji-Rong Wen}.} \bibinfo{year}{2020}\natexlab{}.
\newblock \showarticletitle{{W}asserstein Distance Regularized Sequence
  Representation for Text Matching in Asymmetrical Domains}. In
  \bibinfo{booktitle}{\emph{Proceedings of the 2020 Conference on Empirical
  Methods in Natural Language Processing (EMNLP)}}.
  \bibinfo{publisher}{Association for Computational Linguistics},
  \bibinfo{address}{Online}, \bibinfo{pages}{2985--2994}.
\newblock
\urldef\tempurl%
\url{https://aclanthology.org/2020.emnlp-main.239}
\showURL{%
\tempurl}


\bibitem[Yu et~al\mbox{.}(2022)]%
        {yu2022distribution}
\bibfield{author}{\bibinfo{person}{Weijie Yu}, \bibinfo{person}{Chen Xu},
  \bibinfo{person}{Jun Xu}, \bibinfo{person}{Liang Pang}, {and}
  \bibinfo{person}{Ji-Rong Wen}.} \bibinfo{year}{2022}\natexlab{}.
\newblock \showarticletitle{Distribution Distance Regularized Sequence
  Representation for Text Matching in Asymmetrical Domains}.
\newblock \bibinfo{journal}{\emph{IEEE/ACM Transactions on Audio, Speech, and
  Language Processing}}  \bibinfo{volume}{30} (\bibinfo{year}{2022}),
  \bibinfo{pages}{721--733}.
\newblock


\bibitem[Zeng et~al\mbox{.}(2005)]%
        {zeng2005knowledge}
\bibfield{author}{\bibinfo{person}{Yiming Zeng}, \bibinfo{person}{Ruili Wang},
  \bibinfo{person}{John Zeleznikow}, {and} \bibinfo{person}{Elizabeth~A.
  Kemp}.} \bibinfo{year}{2005}\natexlab{}.
\newblock \showarticletitle{Knowledge Representation for the Intelligent Legal
  Case Retrieval}. In \bibinfo{booktitle}{\emph{Knowledge-Based Intelligent
  Information and Engineering Systems, 9th International Conference, {KES}
  2005, Melbourne, Australia, September 14-16, 2005, Proceedings, Part {I}}}
  \emph{(\bibinfo{series}{Lecture Notes in Computer Science},
  Vol.~\bibinfo{volume}{3681})}. \bibinfo{publisher}{Springer},
  \bibinfo{pages}{339--345}.
\newblock
\urldef\tempurl%
\url{https://doi.org/10.1007/11552413\_49}
\showURL{%
\tempurl}


\bibitem[Zhang et~al\mbox{.}(2020)]%
        {zhang2020pegasus}
\bibfield{author}{\bibinfo{person}{Jingqing Zhang}, \bibinfo{person}{Yao Zhao},
  \bibinfo{person}{Mohammad Saleh}, {and} \bibinfo{person}{Peter~J. Liu}.}
  \bibinfo{year}{2020}\natexlab{}.
\newblock \showarticletitle{{PEGASUS:} Pre-training with Extracted
  Gap-sentences for Abstractive Summarization}. In
  \bibinfo{booktitle}{\emph{Proceedings of the 37th International Conference on
  Machine Learning, {ICML} 2020, 13-18 July 2020, Virtual Event}}
  \emph{(\bibinfo{series}{Proceedings of Machine Learning Research},
  Vol.~\bibinfo{volume}{119})}. \bibinfo{publisher}{{PMLR}},
  \bibinfo{pages}{11328--11339}.
\newblock
\urldef\tempurl%
\url{http://proceedings.mlr.press/v119/zhang20ae.html}
\showURL{%
\tempurl}


\bibitem[Zhao and Vydiswaran(2021)]%
        {zhao2021lirex}
\bibfield{author}{\bibinfo{person}{Xinyan Zhao} {and}
  \bibinfo{person}{V.~G.~Vinod Vydiswaran}.} \bibinfo{year}{2021}\natexlab{}.
\newblock \showarticletitle{LIREx: Augmenting Language Inference with Relevant
  Explanations}. In \bibinfo{booktitle}{\emph{Thirty-Fifth {AAAI} Conference on
  Artificial Intelligence, {AAAI} 2021, Virtual Event, February 2-9, 2021}}.
  \bibinfo{publisher}{{AAAI} Press}, \bibinfo{pages}{14532--14539}.
\newblock
\urldef\tempurl%
\url{https://ojs.aaai.org/index.php/AAAI/article/view/17708}
\showURL{%
\tempurl}


\bibitem[Zhong et~al\mbox{.}(2020)]%
        {zhong2020iteratively}
\bibfield{author}{\bibinfo{person}{Haoxi Zhong}, \bibinfo{person}{Yuzhong
  Wang}, \bibinfo{person}{Cunchao Tu}, \bibinfo{person}{Tianyang Zhang},
  \bibinfo{person}{Zhiyuan Liu}, {and} \bibinfo{person}{Maosong Sun}.}
  \bibinfo{year}{2020}\natexlab{}.
\newblock \showarticletitle{Iteratively Questioning and Answering for
  Interpretable Legal Judgment Prediction}. In \bibinfo{booktitle}{\emph{The
  Thirty-Fourth {AAAI} Conference on Artificial Intelligence, {AAAI} 2020, New
  York, NY, USA, February 7-12, 2020}}. \bibinfo{publisher}{{AAAI} Press},
  \bibinfo{pages}{1250--1257}.
\newblock
\urldef\tempurl%
\url{https://aaai.org/ojs/index.php/AAAI/article/view/5479}
\showURL{%
\tempurl}


\end{thebibliography}

\end{document}